\documentclass[9pt,twocolumn]{article}

\usepackage[margin=0.5 in]{geometry}
\usepackage{amsmath}
\usepackage{amsfonts,amssymb}
\usepackage{graphicx}
\usepackage[T1]{fontenc}
\usepackage[latin1]{inputenc}
\usepackage{caption}
\usepackage{cite}
\usepackage{physics}

\usepackage{xcolor}
\usepackage[pdfborder={0 0 0.1}]{hyperref}
\hypersetup{
	colorlinks=false
}

\title{Reverberant Elastography for the Elastic Characterization of Anisotropic Tissues}

\author{Luis A. Alem{\'a}n-Casta\~{n}eda$^{1,2,*}$ ,
	Fernando Zvietcovich$^{3,4}$ ,
	Kevin J. Parker$^{3,**}$ ,} 
	
\date{
		\small \textit{
			$^1$The Institute of Optics, University of Rochester, Rochester, NY, 14627, USA.\\%
			$^2$Aix Marseille Univ, CNRS, Centrale Marseille, Institut Fresnel, Marseille, France.\\
			$^3$Department of Electrical Computer Engineering, University of Rochester, Rochester, NY, 14627, USA.\\
			$^4$Department of Biomedical Engineering, University of Houston, Houston, TX, 77204,USA\\	
			$^*$lalemanc@ur.rochester.edu, $^{**}$fzvietco@ur.rochester.edu, $^{***}$kevin.parker@rochester.edu\\
			[2ex]%
			(Dated: \today)}
	}

\begin{document}

\twocolumn[
\begin{@twocolumnfalse}
	\maketitle
	\begin{abstract}
	We derive closed-form solutions for reverberant elastography in anisotropic elastic media by adapting the framework used in electromagnetic theory to treat transverse isotropic materials. Different sample-setup geometries are analyzed, highlighting their relevance for both optical coherence elastography (OCE) and ultrasound elastography (USE). Numerical simulations using finite elements are used to validate the proposed solutions in practical cases. OCE experiments are conducted in \emph{ex vivo} chicken muscle samples for the characterization of in-plane and out-of-plane shear modulus assuming a transverse isotropic elastic model. Additionally, we obtained a generalized geometry-independent solution for the isotropic media case, thus unifying previous results for reverberant elastography.
	\end{abstract}
\end{@twocolumnfalse}
]

\section{Introduction}
\label{sect:intro}  

In the field of wave-based elastography, shear waves are used to characterize biomechanical properties of tissues \cite{Parker_2010}. While isotropy is a common assumption, tissues (e.g. muscle, heart, tendon, kidney, and possibly the brain) have an underlying principal direction of structures. Such principal direction is also known as the axis-of-symmetry in a transverse isotropic model of elasticity in solids \cite{Feng_2013,LEVINSON_1987}, or crystal/optic axis for electromagnetic wave propagation in anisotropic crystals \cite{Yariv:book,Aleman2016}. Therefore, the study of anisotropy of tissues in elastography is important and continues to be an emerging field.

Historically, contributions in the measurement of tissue anisotropy have been made for transient mechanical wave propagation in ultrasound elastography (USE) \cite{Gennisson_2010, Royer_2011, Wang_2013}, magnetic resonance elastography (MRE) \cite{Schmidt_2016, Chatelin_2016}, and optical coherence elastography (OCE) \cite{Singh_2016, Singh_2019}. Recently, developments in reverberant elastography have been conducted in USE \cite{Parker_2017,Ormachea_2018,Ormachea_2019} and OCE \cite{Zvietcovich_2019}. A  reverberant shear wave (RSW) field is a limiting case of a statistically uniform distribution of plane shear waves propagating in all directions within a 3D elastic medium. Although reverberant elastography has been proven to be very effective in the biomechanical characterization of tissues with complex boundary conditions \cite{Zvietcovich_2019}, and highly attenuating media \cite{Ormachea_2019}, the theoretical derivation still relies on the assumption of an isotropic media. 

In this paper, we present, for the first time, closed-form solutions to the case of RSW in anisotropic media using key concepts in the analysis of anisotropic crystals with electromagnetic waves. We derive analytical expressions to the complex autocorrelation of RSW fields in materials exhibiting a transverse isotropic model of elasticity for variable directions of: (1) the material's axis-of-symmetry, (2) the motion measurement vector direction (sensor), and (3) the complex autocorrelation function. Moreover, we develop a general solution for the isotropic model which includes the previous specific solutions derived in \cite{Zvietcovich_2019}. Analytical results are compared with finite element simulations for further validation. Finally, experimental results in chicken tibialis muscle are conducted using an optical coherence tomography (OCT) acquisition system for the characterization of degree of anisotropy using RSW fields and the proposed analytical solutions.

A different approach to random waves in media is passive elastography, \cite{Catheline_2008,Brum_2008,Gallot_2011,Benech_2013} also known as time reversal elastography. This is a fundamentally separate method: the autocorrelation used in RSW is a complex autocorrelation in both time and space derived from the limiting case of a distribution of waves across all directions, rather than a real autocorrelation only in time. In both cases, passive and RSW elastography, anisotropy has not been considered before.

The organization of this paper is as follows. In Section 2, we recall the theory behind electromagnetic waves in anisotropic media and its direct extension to mechanical shear waves in the reverberant case. In Section 3, the different combinations of shear wave polarizations, the material's  axis-of-symmetry, and sensor directivity are examined, leading to a general treatment of the complex autocorrelation of RSW fields and the estimators that can characterize the anisotropy of tissues. In Section 4, numerical simulation results using finite elements are compared to the analytical equations for validation. In Section 5, OCE experiments are conducted in \emph{ex vivo} chicken muscle samples for the characterization of the shear modulus along the plane-of-isotropy (in-plane) and in the transverse plane parallel to the axis-of-symmetry (out-of-plane), assuming a transverse isotropic elastic model.  Finally, in Section 6, we summarize the contributions of this paper to the field of reverberant elastography and, more generally, the elastography of anisotropic tissues.

\section{Electromagnetic waves in anisotropic media}

\subsection{Introduction}		

The behavior and propagation of electromagnetic waves, as well as  mechanical waves, differs strongly from isotropic to anisotropic materials. In isotropic media, the wave encounters the same response from the material, no matter its propagation and polarization (oscillation or perturbation) directions, resulting in a homogeneous and singular speed of propagation. However, in anisotropic media the response will depend on the direction of the perturbation, which is linked to the propagation direction in the case of shear waves. Hence, the  propagation speed or effective optical index  perceived by the wave will vary within a range depending on its characteristics.

The treatment of light in anisotropic crystals has long been a subject of interest, and modern theories include a formal dielectric tensor and an ellipsoid of wave normals \cite{Jenkins:book,Born:book,Yariv:book,Aleman2016}. In such crystals, a given plane transversal wave can be decomposed in two eigenmodes of propagation, generally called  in uniaxial materials \textit{ordinary} and \textit{extraordinary}. These have orthogonal polarization states, however not necessarily the same speed of propagation. We use plane waves since any field can be expressed using plane-wave decomposition and because they are  compatible with the reverberant studies done previously \cite{Parker_2017,Ormachea_2018,Ormachea_2019,Zvietcovich_2019}. The following approach concerns electromagnetic waves, and it will be extended directly to mechanical shear waves in Section 3. 

\subsection{Theory}	

We will assume an homogeneous and non-magnetic (or at least magnetically isotropic) medium without free charges or currents. Given these assumptions, we can focus  only on the electric field $\mathbf{E}=\mathbf{E_0}e^{i(\mathbf{k}\cdot \mathbf{r}-\omega t)}$, 
where $\mathbf{k}$ is the wave vector, $\mathbf{r}$ is the 3D position vector, and $\omega$ its frequency. For light, we have that $\mathbf{k}=\frac{\omega n_{\text{eff}}}{c}\mathbf{\hat{g}}$, in which $c$ is the speed of light in vacuum, $n_{\text{eff}}$ is the effective refractive index perceived by the wave inside the medium, and $\mathbf{\hat{g}}$ is the unitary wave vector direction. Besides Maxwell's equations, the constitutive relations describe how media responds to electromagnetic fields, in our case the electric field produces an electric displacement field inside the material of $\mathbf{D}=\boldsymbol{\epsilon}\cdot\mathbf{E}$, 
where $\boldsymbol{\epsilon}$ is the dielectric tensor, a second order tensor. As opposed to isotropic materials, $\mathbf{D}\nparallel\mathbf{E}$, which leads to a \textit{walk-off angle}\cite{Born:book,Yariv:book} between the Poynting vector (energy propagation direction) and the wave vector (phase acquisition direction).

In general there exists a coordinate system such that the dielectric tensor becomes a diagonal matrix where the entries are the principal dielectric responses:
\begin{equation}
\boldsymbol{\epsilon}=\begin{pmatrix}
\epsilon_x &0 &0 \\
0& \epsilon_y & 0 \\
0& 0& \epsilon_z 
\end{pmatrix}.
\end{equation}
Another expression for non-magnetic media is given in terms of the principal optical indices and the electric permittivity in a vacuum, $\epsilon_0$, using $n_i^2=\epsilon_i/\epsilon_0$, so the tensor becomes $\epsilon_o\boldsymbol{n}=\boldsymbol{\epsilon}$, where $n_i$ stands for the corresponding principal optic indexes.

Both principal directions and its values depend on the structure of the  media, and so dielectric materials can be grouped in three different categories:
\begin{itemize}
	\item \textit{Isotropic,} where $\epsilon_x=\epsilon_y=\epsilon_z$, so $\boldsymbol{\epsilon}$ can be reduced to a scalar and $\mathbf{D}\parallel\mathbf{E}$. 
	
	\item \textit{Uniaxial birefringent},  where two principal dielectric responses are equal, meaning that there is a plane in which all the directions of perturbation are equivalent. In literature, the two terms that are repeated correspond to the ordinary index $n_o$, and the extraordinary index $n_e$. For these materials there is one unique propagation direction in which the optical index is independent of porlarization, hence the name uniaxial. This direction is referred to as the crystal axis direction\cite{Aleman2016} (we refrain from the term \textit{optic axis}\cite{Yariv:book} to avoid confusion with the system's \textit{optical axis}), which we indicate as $\mathbf{A}$. For example, if the crystal axis were in the $\mathbf{z}$ direction, then the coefficients would be $\epsilon_x=\epsilon_y=\epsilon_0n_o^2$ and $\epsilon_z=\epsilon_0n_e^2$. 
	
	\item \textit{Biaxial birefringent}, where $\epsilon_x\neq\epsilon_y\neq \epsilon_z$.  
	Here there are two directions of propagation in which the optical index is independent of the polarization, hence the name \textit{biaxial}.
\end{itemize}
Nevertheless the experiment's geometry doesn't generally correspond to this very specific coordinate system in which $\boldsymbol{\epsilon}$ is diagonal, and we are interested in what happens when light does not oscillate/propagate in any of the principal directions.
To address this condition, in both an analytical and graphical way, we need to use the wave equation, which in the $k$-domain (or using our plane-wave assumption) is 
\begin{equation}
\mathbf{k}\times\left(\mathbf{k}\times\mathbf{E}\right)=-\omega^2 \mu \boldsymbol{\epsilon}\mathbf{E},
\end{equation}
where $\mu$ is the magnetic permeability of the material. Then, using  $c^2\approx1/(\mu\epsilon_0)$ since we are assuming non-magnetic materials\cite{Born:book,Yariv:book}, the equation is reduced to a homogeneous system
\begin{equation} \label{eq:wave_eq__plane}
\left((\mathbf{\hat{g}}\mathbf{\hat{g}})-\mathbb{I}+\frac{1}{n_{\text{eff}}^2}\mathbf{n}\right)\mathbf{E}=0,
\end{equation}
where $\mathbb{I}$ is the identity matrix, and $(\mathbf{\hat{g}}\mathbf{\hat{g}})$ is the dyadic product (i.e. the tensor whose entries are of the form $g_ig_j$, where $g_{i}$ is the components of the normalized wave vector).  Note that the determinant of Eq. (\ref{eq:wave_eq__plane}) must vanish in order to obtain non-trivial solutions. Hence we obtain 
\begin{equation}\label{eq:normal_surfaces}
\mathcal{G}:~~~\begin{vmatrix}
\frac{n^2_x}{n^2_{\text{eff}}}-(1-g_x^2)  & g_xg_y & g_xg_z\\
g_yg_x & \frac{n^2_y}{n^2_{\text{eff}}}-(1-g_y^2)  & g_yg_z\\
g_zg_x & g_zg_y& \frac{n^2_z}{n^2_{\text{eff}}}-(1-g_z^2) \\
\end{vmatrix}=0.
\end{equation}

\begin{figure}[!ht]
	\centering
	\includegraphics[width=0.9\linewidth]{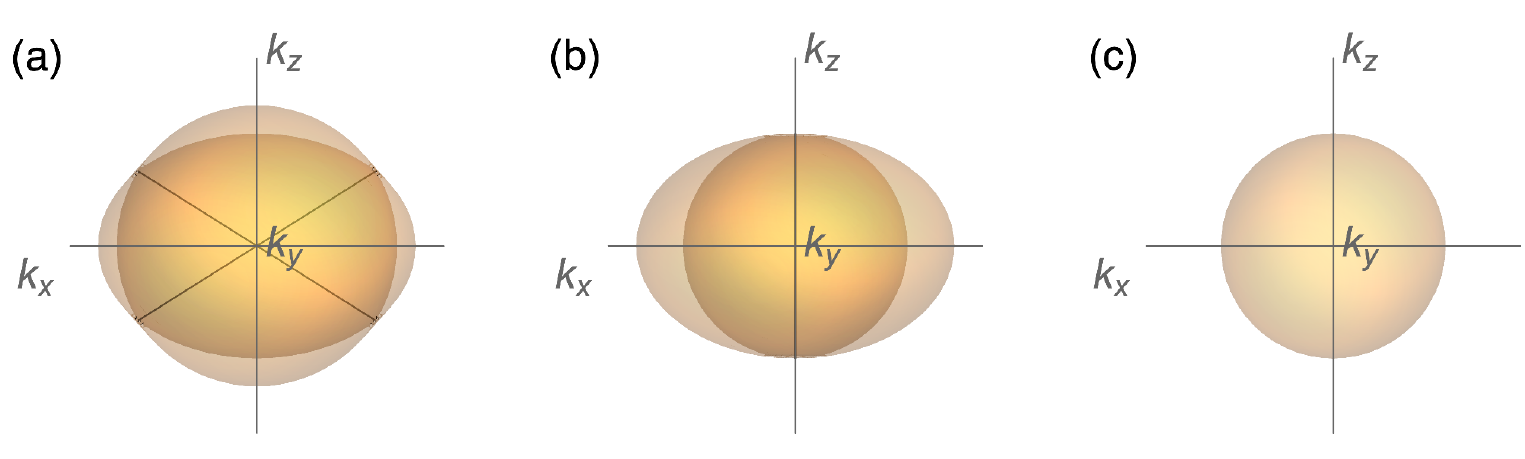}
	\caption{Hypothetical normal-surfaces with their respective crystal axes shown as thick black lines.  (a) Biaxial case, here $n_x<n_y<n_z$, so the crystal axes lie in the $x$-$z$ plane. The surfaces cannot be expressed in terms of simple geometrical objects. (b) Uniaxial case with $n_o=n_x=n_y$ and $n_x<n_z=n_e$. One of the surfaces is always a sphere, while the other is an ellipsoid that touches the sphere along the crystal axis. (c) Isotropic case, for which both shells become one single sphere and there is no definite crystal axis.}
	\label{fig:normal_surface}
\end{figure}

The surfaces defined by this equation consist of two shells in $k$-space,  also called  normal-surfaces, which have a nice interpretation: they are the surfaces made by all the eigenvectors of the material, meaning that in any direction there are two eigenmodes of wave propagation that have different wave vector magnitudes and have orthogonal polarizations with respect to each other. In other words, a plane wave propagating inside the material in a given direction will be decomposed in to two parallel-propagating plane waves which perceive, in general, different effective optical indices. Commonly, these two shells have four points in common (biaxial  materials have four points, while uniaxial only two), and the lines that pass through them and the origin define the crystal axes previously discussed, see Figure \ref{fig:normal_surface}.

To determine the polarization (oscillation or perturbation) direction of each propagating plane wave eigenmode, one needs to solve the full eigenvalue/eigenvector problem, i.e. solve the wave equation given by Eq. (\ref{eq:wave_eq__plane}).
This problem can also be handled via the Fresnel equations \cite{Born:book}. In the following subsections isotropic and uniaxial scenarios are discussed, and their polarization states are described. The biaxial case is far more cumbersome, nevertheless, as expected,the eigenmodes are orthogonally polarized, i.e. $\mathbf{D_1}\cdot\mathbf{D_2}=0$, where the subscripts 1 and 2 are their corresponding labels.

\subsubsection{Isotropic media}
As shown in Figure \ref{fig:normal_surface}c, in isotropic materials the normal surface becomes a single sphere centered at the origin, so in each direction of propagation the  wave vector will have the same magnitude, i.e., perceive  the same effective index. 
Additionally, since the eigenmode problem is degenerate, any plane wave with a given polarization can be regarded as an eigenmode of propagation.

\subsubsection{Uniaxial media}
Uniaxial  birefringent materials (analogous to cornea and muscles for mechanical waves) can be described by a crystal axis direction $\mathbf{\hat{A}}$ and  two optical indices: the ordinary $n_o$ and the extraordinary $n_e$ indices. It follows that Eq. (\ref{eq:normal_surfaces}) simplifies to
\begin{equation}\label{eq:unniaxial_normal}
\mathcal{G}:~~~ \left(\frac{n^2_{\text{eff}}}{n_o^2}-1\right)\left(\frac{n^2_{\text{eff}}\sin^2\psi}{n_e^2}+\frac{n^2_{\text{eff}}\cos^2\psi}{n_o^2}-1\right)=0,
\end{equation}
where $\psi$ is the angle between the wave vector and the crystal axis, i.e. $\cos\psi=\mathbf{\hat{g}}\cdot\mathbf{\hat{A}}$. The normal-surfaces correspond to a sphere (first term) and an ellipsoid (second term), both centered at the origin and that touch along the crystal axis direction. Note that the ellipsoid is symmetric with respect to the crystal axis.  According to Eq. (\ref{eq:unniaxial_normal}), one of the two eigenmodes of wave-propagation perceives the same effective index no matter its direction of propagation, while for the other eigenmode, $n_2$, it depends on the angle of the wave vector with respect to the crystal axis, varying within the two extremae values $n_o$ and $n_e$. Explicitly, 
defining $k_{\text{eff}}=2\pi n_{\text{eff}}/\lambda$, we have
\begin{equation}\label{eq:k_dependnecy}
\begin{aligned}
k_{\text{eff},1}&=k_o, &  k_{\text{eff},2}&=\frac{k_ok_e}{\sqrt{k_o^2\sin^2\psi +k_e^2\cos^2\psi}},
\end{aligned}
\end{equation}
which is shown for two hypothetical $k_o$ and $k_e$ in Figure \ref{fig:normal_surface}b.

As mentioned previously, the effective index will vary depending on the wave propagation direction, and also on its polarization state, which can be obtained from solving the eigenvalue/eigenvector problem in Eq. (\ref{eq:wave_eq__plane}). The resulting polarization directions (normalized)\cite{Yariv:book} are
\begin{equation}\label{eq:optical_polarizations}
\begin{aligned}
\mathbf{\hat{D}}_1&=\frac{\mathbf{\hat{g}}\times\mathbf{\hat{A}}}{|\mathbf{\hat{g}}\times\mathbf{\hat{A}}|}=\frac{\mathbf{\hat{g}}\times\mathbf{\hat{A}}}{\sin\psi}, \\ \mathbf{\hat{D}}_2&=\frac{(\mathbf{\hat{g}}\times\mathbf{\hat{A}})\times\mathbf{\hat{g}}}{|\mathbf{\hat{g}}\times(\mathbf{\hat{g}}\times\mathbf{\hat{A}})|}=
\frac{\mathbf{\hat{A}}-\mathbf{\hat{g}}\cos\psi }{\sin\psi},
\end{aligned}
\end{equation} 
meaning that an ordinary mode does not have any component along the crystal axis direction.

\section{Reverberant elastography in anisotropic media}

\subsection{Introduction}

The generalization of the case involving mechanical waves is far more complicated than the electromagnetic case (Section 2). While in electromagnetism waves are only transversal, in the mechanical case, the elastic media support the propagation of three types of waves: two shear waves with orthogonal and transversal motion polarization, and one compression wave with longitudinal motion polarization. Furthermore, the role of the $3\times3$ electromagnetic tensor is now played by the stiffness tensor $\mathbf{c}$, a $3\times3\times3\times3$ tensor. Fortunately instead of 81 coefficients, given symmetry and energy conservation conditions, $\mathbf{c}$ only has 21 independent elements\cite{Graff:book} -- compared with $\boldsymbol{\epsilon}$ that has three independent elements.

In this section, the reverberant theory is extended to anisotropic materials, specifically to uniaxial birefringent media which, in elastic solids, is equivalent to the transverse isotropic model \cite{Feng_2013}. The expressions for wave-number $\mathbf{k}$ (equivalent to effective index for electromagnetic waves), and the motion direction (polarization states for electromagnetic waves) of the mechanical wave perturbation need to be defined from Section 2 since they transit from electromagnetic to mechanical shear waves. We are extending the transversal wave dynamics of light into elastic bodies, ignoring completely the compression waves, which in any case propagate at much higher speeds and are not considered in this paper. In Section 3.2, we revisit the isotropic reverberant case providing a generalization of equations provided in previous works\cite{Parker_2017,Ormachea_2018,Ormachea_2019,Zvietcovich_2019} for specific cases, and finalizing with the derivation for the anisotropic case in Section 3.3.

\subsection{Isotropic media}
A spatio-temporal particle velocity (motion) reverberant field is defined as $\mathbf{V}(\mathbf{r},t)$, where $\boldsymbol{r}$ represents the 3D position vector and $t$ is time. This field 
is the superposition of all possible plane shear waves traveling in random directions with the same wave-number, $k=|\mathbf{k}|$, and frequency, $\omega_0$, 
\begin{equation}
\mathbf{V}(\mathbf{r},t)=\sum_{q,l}\mathbf{\hat{V}}_{ql} v_{ql}e^{i\left(k\mathbf{\hat{g}}_q\cdot\boldsymbol{r}-\omega_0t\right)}.
\end{equation}
The subscript $q$ specifies a realization of $\mathbf{\hat{g}}_q$, a random unit vector indicating the direction of wave propagation, and the index $l$ indicates a realization of $\mathbf{\hat{V}}_{ql}$, the random vector describing the direction of perturbation (particle velocity for mechanic waves, corresponding to polarization of light for the field $\mathbf{\hat{D}}$). Since we are dealing with transversal waves, $\mathbf{\hat{V}}_{ql}\cdot \mathbf{\hat{g}}_q=0$. Lastly, $v_{ql}$ is an independent, identically-distributed random variable describing the magnitude of the particle velocity within a realization. The summation over $q$ is understood to be taken over the $4\pi$ solid angle, while over $l$ it is taken over a $2\pi$ angle within a disk perpendicular to the wave direction given by $\mathbf{\hat{g}}_q$.

Here we proceed differently than in previous works \cite{Parker_2017,Ormachea_2018,Ormachea_2019,Zvietcovich_2019} using the fact that any oscillation can be decomposed in a vector basis consisting of two directions orthogonal to the wave propagation, $\mathbf{\hat{g}}$. Thus our sampling consists of independent realizations of these two directions, instead of sampling overall possible directions of oscillation.
These two approaches are equivalent and arrive at the same expressions, nevertheless we opt for the decomposition method since it can be extended directly to tackle the anisotropic problem. 

\begin{figure}[!ht]
	\centering
	\includegraphics[width=0.9\linewidth]{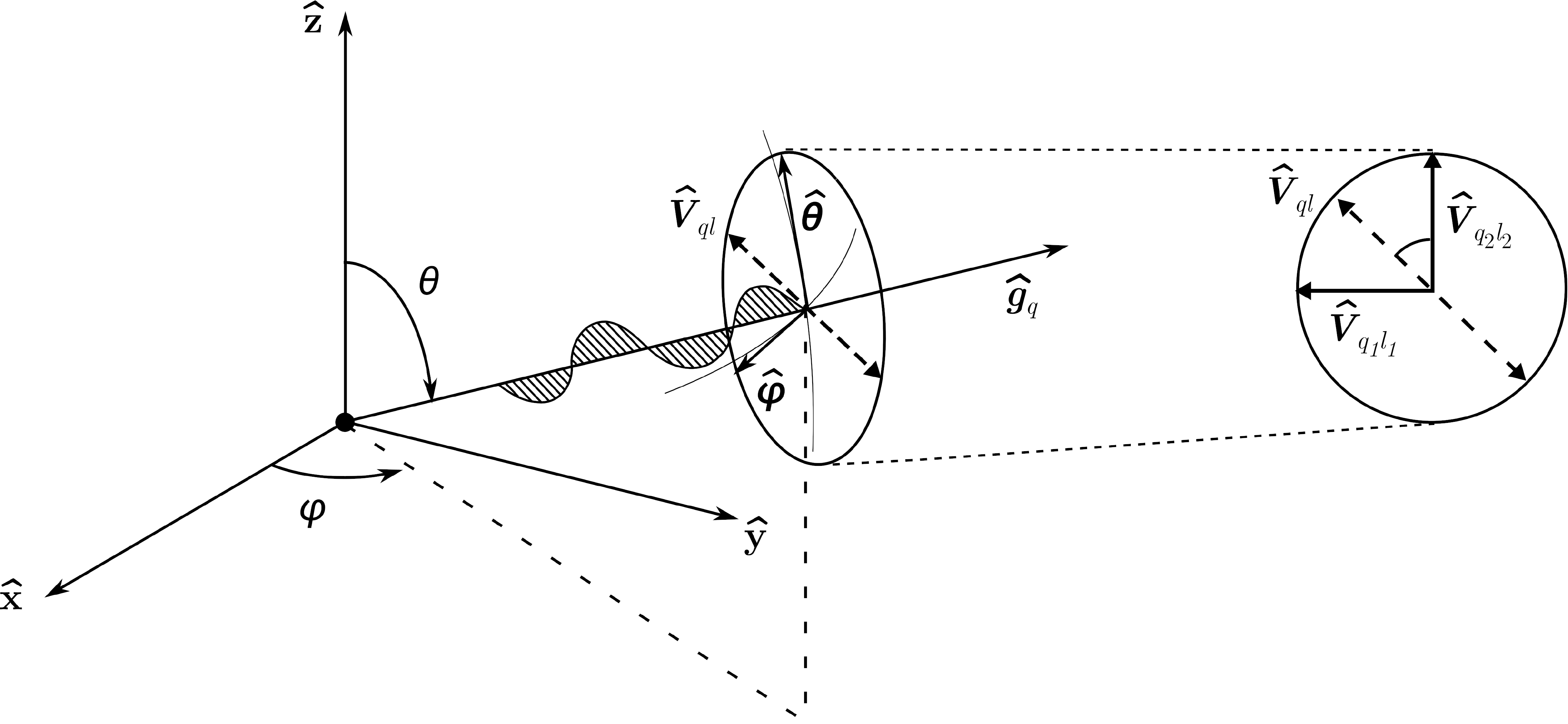}
	\caption{Mode decomposition for any shear wave with direction given by  $\mathbf{\hat{g}}_q$  (radial direction, defined by the angles $\theta$ and $\varphi$). Any perturbation direction  $\mathbf{\hat{V}}_{ql}$, since it is transversal, can be expressed in terms of its $\boldsymbol{\hat{\theta}}$ and $\boldsymbol{\hat{\varphi}}$ components.
		Therefore, instead of sampling randomly this perturbation direction and obtaining the projections, we sample each mode independently, i.e. sample $\mathbf{\hat{V}}_{q_1,l_1}$ and $\mathbf{\hat{V}}_{q_2,l_2}$.  }
	\label{fig:geometry}
\end{figure}

Let us use spherical coordinates to express the direction of wave propagation. For simplicity we choose the following basis, note the resemblance to Eq. (\ref{eq:optical_polarizations}),
\begin{equation}\label{eq:decmposition_iso}
\begin{aligned}
\mathbf{\hat{V}}_{1}&=\frac{\mathbf{\hat{g}}\times\mathbf{\hat{z}}}{\sqrt{1-(\mathbf{\hat{g}}\cdot\mathbf{\hat{z}})^2}}=\boldsymbol{\hat{\varphi}}, \\ \mathbf{\hat{V}}_2&=\frac{(\mathbf{\hat{g}}\times\mathbf{\hat{z}})\times\mathbf{\hat{g}}}{\sqrt{1-(\mathbf{\hat{g}}\cdot\mathbf{\hat{z}})^2}}=\boldsymbol{\hat{\theta}},
\end{aligned}
\end{equation}
where $\boldsymbol{\hat{\theta}}=\cos(\theta)\cos(\varphi)\mathbf{\hat{x}}+\cos(\theta)\sin(\varphi)\mathbf{\hat{y}}-\sin(\theta)\mathbf{\hat{z}}$ and $\boldsymbol{\hat{\varphi}}=\cos(\varphi)\mathbf{\hat{y}}-\sin(\varphi)\mathbf{\hat{x}}$ are the unit vectors in the polar and azimuthal directions at ($\theta$,$\varphi$), respectively, and consequently $\mathbf{\hat{x}},~\mathbf{\hat{y}}$ and $\mathbf{\hat{z}}$ are the unitary Cartesian coordinate vectors, see Figure \ref{fig:geometry}. Therefore we have
\begin{equation}\begin{split}
\mathbf{V}(\mathbf{r},t)=\sum_{q_1,l_1}\mathbf{\hat{V}}_{q_1,l_1} v_{q_1,l_1}e^{i\left(k_1\mathbf{\hat{g}}_{q_1}\cdot\boldsymbol{r}-\omega_0t\right)}+ \\
\sum_{q_2,l_2}\mathbf{\hat{V}}_{q_2l_2} v_{q_2l_2}e^{i\left(k_2\mathbf{\hat{g}}_{q_2}\cdot\boldsymbol{r}-\omega_0t\right)},
\end{split}\end{equation}
where both contributions come from independent realizations.

Given that ultrasound and OCT systems typically measure the particle velocity in one direction, which we denote as the sensor axis, $\mathbf{\hat{e}_s}$, we will project this resulting particle velocity, $	V_s(\mathbf{r},t)=\mathbf{V}(\mathbf{r},t)\cdot\mathbf{\hat{e}}_s$,  according to the desired geometry, 
\begin{equation}\label{eq:V_projected}\begin{split}
V_s(\mathbf{r},t)=\sum_{q_1,l_1}V_{q_1,l_{1s}} v_{q_1,l_1}e^{i\left(k\mathbf{\hat{g}}_{q_1}\cdot\boldsymbol{r}-\omega_0t\right)} + \\ 
\sum_{q_2,l_2}V_{q_2l_{2s}}v_{q_2l_2}e^{i\left(k\mathbf{\hat{g}}_{q_2}\cdot\boldsymbol{r}-\omega_0t\right)},
\end{split}\end{equation}
where $V_{ql_s}=\mathbf{\hat{V}}_{ql}\cdot\mathbf{\hat{e}}_s$ becomes a scalar random variable. We are interested in the autocorrelation function of Eq. (\ref{eq:V_projected}) in both space and time, which we denote as $B_{V_sV_s}$, and is defined as 
\begin{equation}\label{eq:full_autocorrelation}
B_{V_sV_s}(\boldsymbol{\Delta r},\Delta t)=\mathbb{E}\left\{V_s(\mathbf{r},t)V_s^*(\mathbf{r}+\boldsymbol{\Delta r},t+\Delta t)\right \}
\end{equation}
where $\mathbb{E}$ represents an ensemble average and the asterisk represents conjugation. Many of the terms correspond to cross terms which will vanish given that they correspond to independent realizations, so Eq. (\ref{eq:full_autocorrelation}) simplifies to
\begin{equation}\begin{split}\label{eq:full_autocorrelation1}
B_{V_sV_s}(\boldsymbol{\Delta r},\Delta t)=\frac{\overline{v^2}}{2}e^{i\omega_0\Delta t}~ \times~~~~~~~~~ \\ \mathbb{E}\left\{  \sum_{q_1,l_1} V_{{q_1l_1}_s}^2 e^{-ik\mathbf{\hat{g}}_{q_1}\cdot\boldsymbol{\Delta r}}+ 
\sum_{q_2,l_2} V_{{q_2l_2}_s}^2 e^{-ik\mathbf{\hat{g}}_{q_2}\cdot\boldsymbol{\Delta r}} \right\},
\end{split}\end{equation}
in which we renamed  the expected value of  the squared velocity of the particle in each direction, i.e. $\expval{v_{q_1l_1}^2}_{q_1l_1}=\expval{v_{q_2l_2}^2}_{q_2l_2}=\overline{v^2}/2$, assuming that each component has half the energy. Note that we could factor Eq. (\ref{eq:full_autocorrelation1}) out given the independence between $v_{q l}$ and $\{\hat{g}_q,V_{ql_s}\}$. In an ideal reverberant field this ensemble average becomes the average over all possible directions of wave propagation (over $4\pi$), specified in spherical coordinates with $(\theta,\varphi)$. Therefore, renaming $B_{V_sV_s}:=B_{\text{iso}}$, we have
\begin{equation}\label{eq:B_isotropic_integral}\begin{split}
B_{iso}(\boldsymbol{\Delta r},\Delta t)=\frac{\overline{v^2}}{8\pi}e^{i\omega_0\Delta t}\int_{0}^{2\pi}\int_{0}^{\pi}\left[V_{1,s}^2(\theta,\varphi) + \right. \\ \left.
V_{2,s}^2(\theta,\varphi)\right]e^{-ik\mathbf{\hat{g}}\cdot\boldsymbol{\Delta r}}\sin\theta d\theta d\varphi.
\end{split}\end{equation} 

To solve this integral, we choose the direction of correlation that results in the greatest simplification, i.e. along the $z$-axis, $\boldsymbol{\Delta r}=(\Delta z) \boldsymbol{\hat{z}}$, so
\begin{equation}\label{eq:correlationdirection}
\mathbf{\hat{g}}\cdot(\Delta z)\mathbf{\hat{z}}=k\Delta z \cos\theta.
\end{equation}
We must set the direction along which the particle velocity will be measured (also called sensor axis), and, given the symmetry around the $z$-axis, we choose it to be somewhere along the $xz$ plane, so $\mathbf{\hat{e}}_s=\cos\theta_s \mathbf{\hat{z}}+\sin\theta_s \mathbf{\hat{x}}$, where $\theta_s$ is the angle of the sensor with respect the $z$ axis:
\begin{equation}\label{eq:velocitydecomposition}
\begin{aligned}
V_{1,s}(\theta,\varphi)&=-\sin\varphi\sin\theta_s, \\
V_{2,s}(\theta,\varphi)&=\cos\varphi\cos\theta\sin\theta_s-\sin\theta\cos\theta_s.
\end{aligned}
\end{equation} 

\begin{figure}[!ht]
	\centering
	\includegraphics[width=\linewidth]{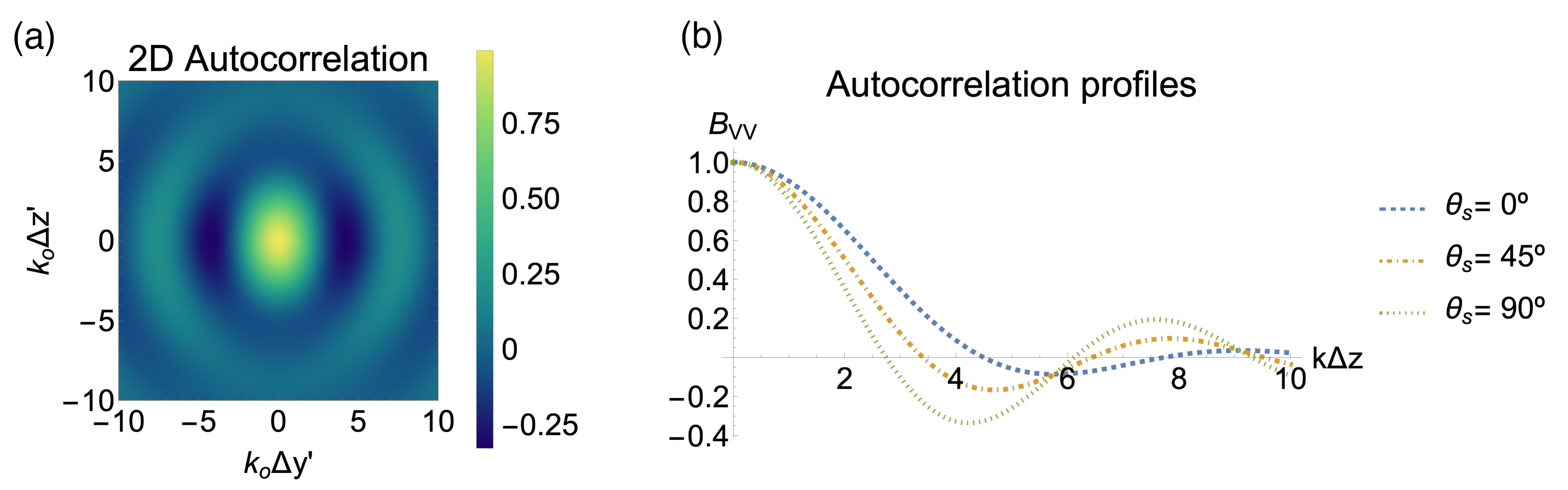}
	\caption{Autocorrelation functions for an isotropic material. The functions are normalized, and the actual maximum at the origin is $1/3$. (a) 2D autocorrelation map. The sensor direction corresponding to the $z'$ axis. (b) 1D profiles for different correlation directions, with an angle $\theta_s$ with respect the sensor axis. Only one half of the plots are shown given that they are symmetric.}
	\label{fig:Bvv_FullIsotropic}
\end{figure}

Note that whenever $\theta_s=0$ the sensor is parallel to the correlation direction, while when $\theta_s=\pi/2$ the sensor and correlation directions become perpendicular. These canonical scenarios are the two cases that have been studied previously \cite{Parker_2017,Ormachea_2018,Ormachea_2019,Zvietcovich_2019}. Substituting Eqs. (\ref{eq:correlationdirection}-\ref{eq:velocitydecomposition}) into Eq. (\ref{eq:B_isotropic_integral}) and solving the integral leads to
\begin{equation}\label{eq:isotropic_result}
\begin{aligned}\begin{split}
B_{\text{iso}}(\Delta z,\Delta t)=\overline{v^2}e^{i\omega_0 \Delta t} \left\{ \frac{\sin^2\theta_s}{2}  \left[ j_0 (k \Delta z)- \right. \right. \\ \left. \left.
\frac{j_1(k \Delta z)}{k \Delta z} \right]+\cos^2\theta_s \frac{j_1(k \Delta z)}{k \Delta z} \right\},
\end{split}\end{aligned}
\end{equation}
where $j_n(x)$ are the spherical Bessel functions of order $n$. Analogously, considering the setup frame in which the sensor is generally fixed, we can define the sensor axis to be the $z'$ axis and interpret $\theta_s$ as the autocorrelation direction angle with respect to $z'$ (sensor axis). Then, note that Eq. (\ref{eq:isotropic_result}) is a linear combination of the two canonical cases: correlation parallel or perpendicular to the sensor reported in \cite{Parker_2017, Zvietcovich_2019}. It follows that the width of the  central region is related to the wave-number, $k$, and so its value can be estimated by fitting the measurements, see Figure \ref{fig:Bvv_FullIsotropic}. 

\subsection{Anisotropic media: uniaxial case}

Unlike the isotropic case, for uniaxial materials we cannot select any two vectors to decompose the oscillation, but instead we have to use the natural decomposition in ordinary and extraordinary modes. Our assumption is that both eigenmodes are equally represented and that each carries half of the energy, given the reverberant chamber condition. Therefore, the reverberant field is given by the summation of ordinary and extraordinary waves,
\begin{equation}\begin{split}
\mathbf{V}(\mathbf{r},t)=\sum_{q_1,l_1}\mathbf{\hat{V}}_{q_1l_1} v_{q_1l_1}e^{i\left(k_1\mathbf{\hat{g}}_{q_1}\cdot\boldsymbol{r}-\omega_0t\right)}+ \\
\sum_{q_2,l_2}\mathbf{\hat{V}}_{q_2l_2} v_{q_2l_2}e^{i\left(k_2\mathbf{\hat{g}}_{q_2}\cdot\boldsymbol{r}-\omega_0t\right)},
\end{split}\end{equation}
where the labels $1$ and $2$ stand for ordinary and extraordinary modes, respectively. Similarly to the isotropic case, both contributions are independent from each other and random, so cross terms  vanish. Consequently, the autocorrelation ends up being the average of both ordinary and extraordinary contributions over $4\pi$, so
\begin{equation}\label{eq:B_general_integral}\begin{split}
B_{\text{aniso}}(\boldsymbol{\Delta r},\Delta t)=\frac{\overline{v^2}}{8\pi}e^{i\omega_0\Delta t}\int_{0}^{2\pi}\int_{0}^{\pi}\left(V_{1,s}^2 e^{-ik_1\mathbf{\hat{g}}\cdot\boldsymbol{\Delta r}} \right. \\ \left.
+V_{2,s}^2 e^{-ik_2\mathbf{\hat{g}}\cdot\boldsymbol{\Delta r}}\right)\sin\theta d\theta d\varphi.
\end{split}\end{equation}

Before proceeding, we need to revisit the corresponding oscillation directions for each eigenmode. Unlike the electromagnetic case, in which the electric field $\mathbf{E}$ may oscillate along any arbitrary direction and the dielectric tensor responds differently to each direction, for mechanical shear waves the stiffness tensor and the stress are defined in planes rather than directions. Let us consider only the shearing dynamics of a transverse isotropic elastic model of a linear-elastic medium and write the corresponding part of the stiffness tensor in the coordinate system which diagonalizes it \cite{Graff:book},
\begin{equation}
\begin{pmatrix}
\gamma_{X'Y'}\\
\gamma_{X'Z'}\\
\gamma_{Y'Z'}
\end{pmatrix}=\begin{pmatrix}
1/G_e & 0 & 0\\
0& 1/G_o &0\\
0 &0& 1/G_o
\end{pmatrix} \begin{pmatrix}
\sigma_{X'Y'}\\
\sigma_{X'Z'}\\
\sigma_{Y'Z'}
\end{pmatrix}.
\end{equation}

Since it is still a $3\times 3$ tensor, the mathematics remain the same as in the electromagnetic case, however, the physical interpretation changes dramatically. Here the eigenvalue corresponding to the extraordinary mode (multiplicity of one) is related to shear deformations along the plane perpendicular to the axis-of-symmetry, $\boldsymbol{\hat{A}}$  ($\boldsymbol{\hat{z}'}$). However, the ordinary eigenvalue is related to components that include this axis. Therefore, the oscillation of each eigenmode propagating along $\mathbf{\hat{g}}$ are swapped with respect to the electromagnetic case, i.e.
\begin{equation}\label{eq:mechanical_polarizations}
\begin{aligned}
\mathbf{\hat{V}}_1&=\frac{(\mathbf{\hat{g}}\times\mathbf{\hat{A}})\times\mathbf{\hat{g}}}{\sqrt{1-(\mathbf{\hat{g}}\cdot\mathbf{\hat{A}})^2}}, & \mathbf{\hat{V}}_2&=\frac{\mathbf{\hat{g}}\times\mathbf{\hat{A}}}{\sqrt{1-(\mathbf{\hat{g}}\cdot\mathbf{\hat{A}})^2}}.
\end{aligned}
\end{equation} 

\begin{figure}[!ht]
	\centering
	\includegraphics[width=1\linewidth]{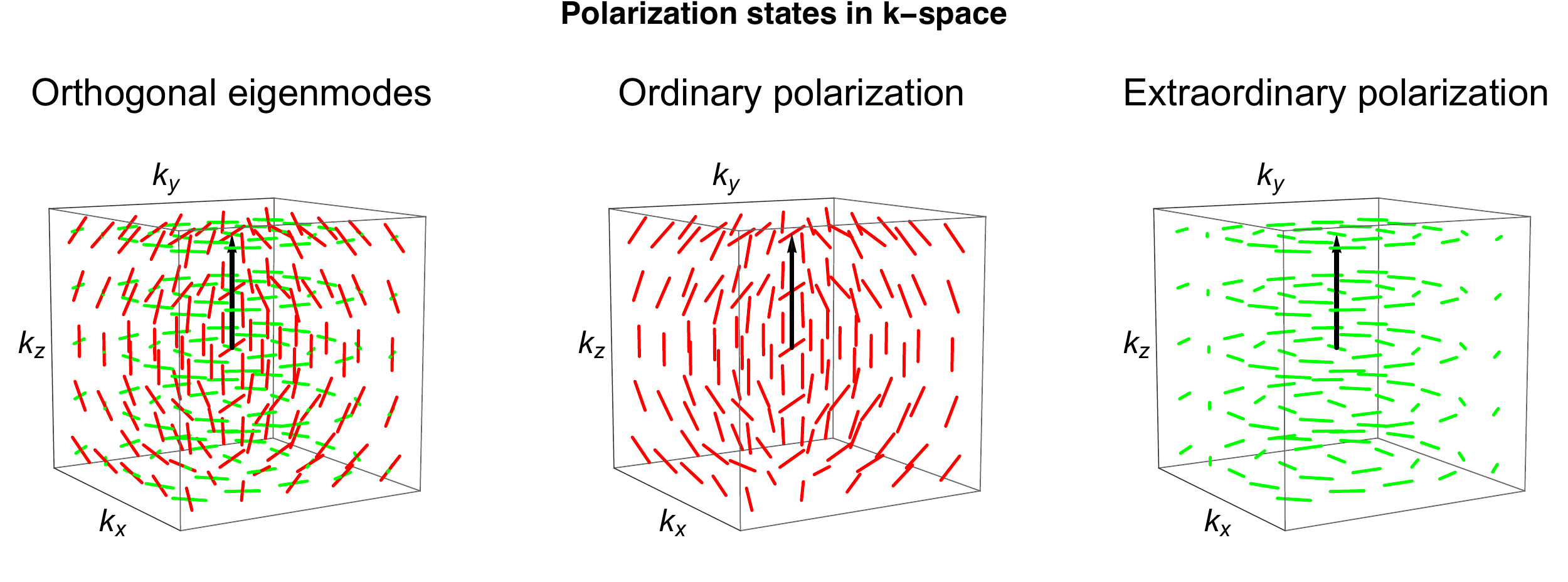}
	\caption{The polarization eigenmodes are shown in k-space, all extraordinary modes do not have any component along the axis-of-symmetry (black arrow). }
	\label{fig:Polarization_Uniaxial}
\end{figure}

Additionally, when considering anisotropy not only do calculations get convoluted, but more cases appear since the axis-of-symmetry (crystal axis in optics) $\mathbf{\hat{A}}$ has to be considered along with the correlation direction and the sensor axis. Nevertheless, there is an immediate conclusion obtained from Eq. (\ref{eq:mechanical_polarizations}): the extraordinary mode oscillation doesn't have any component along the axis-of-symmetry, see Figure \ref{fig:Polarization_Uniaxial}. As a result, whenever the sensor is along $\mathbf{\hat{A}}$, only the ordinary contribution will be measured, and so we expect $k$ to be related only to $k_o$. In the rest of the cases we expect the extraordinary 
contribution to spread the range of values of $k$ within $k_o$ and $k_e$.

In order to solve analytically the integral in Eq. (\ref{eq:B_general_integral}), an approximation must be made about the amount of anisotropy: we assume that it is small. In other words, 
\begin{equation} \label{eq:Delta_e}
|\delta_e|=\left|\frac{k_e^2-k_o^2}{k_e^2}\right|\ll1,
\end{equation}
such that the exponential can be expanded as a Taylor series around $k_o$ and with respect to $\delta_e$, i.e.
\begin{equation}\begin{split}
e^{-ik_2\mathbf{\hat{g}}\cdot\boldsymbol{\Delta r}}\approx e^{-ik_o\mathbf{\hat{g}}\cdot\boldsymbol{\Delta r}}\left(1-i\frac{k_o\delta_e}{2}\left(\mathbf{\hat{g}}\cdot\boldsymbol{\Delta r}\right) \right. \\ \left. \left[1-(\mathbf{\hat{g}}\cdot\mathbf{\hat{A}})^2\right]\right).
\end{split}\end{equation}

Given the expansion, the autocorrelation can be rearranged such that the resulting expressions can be regarded as adding corrections to the isotropic results derived 
in Eq. (\ref{eq:isotropic_result}). Explicitly:
\begin{equation}\begin{split}
B_{\text{aniso}}(\boldsymbol{\Delta r},\Delta t;k_o,k_e)=B_{\text{iso}}(\boldsymbol{\Delta r},\Delta t;k_o)+\\
\delta B(\boldsymbol{\Delta r},\Delta t;k_o,k_e),
\end{split}\end{equation}
where the anisotropic correction, assuming $\mathbf{\hat{e}_s}$ as sensor axis, becomes
\begin{equation}\label{eq:anisotropic_correction}\begin{split}
\delta B=-i\frac{k_o\delta_e}{2}\frac{\overline{v^2}}{8\pi}e^{i\omega_0\Delta t}\int_{0}^{2\pi}\int_{0}^{\pi}\left(\mathbf{\hat{g}}_{q}\cdot\boldsymbol{\Delta r}\right) \\
(\mathbf{\hat{e}_s}\cdot [\mathbf{\hat{g}}\times \mathbf{\hat{A}}])^2 e^{-ik_2\mathbf{\hat{g}}_{q}\cdot\boldsymbol{\Delta r}} \sin\theta d\theta d\varphi,
\end{split}\end{equation}
and in which the explicit dependency of $\delta B$ with respect $\mathbf{\Delta r}$ and $\Delta t$ was dropped. 

There are several studies characterizing anisotropic samples such as muscles \cite{LEVINSON_1987,Wang_2013} and tendons \cite{Brum_2014,Aubry2013}.
Although assuming small anisotropy is acceptable in many optical materials\cite{Yariv:book}, for mechanical waves it may not be, e.g. muscles with weight loads. In these mechanical cases, it may be safer to define $k_m=(k_o+k_e)/2$ and $k_d=(k_e-k_o)/2$, so the expansion can be done around $k_m$ and with respect to the relative anisotropy $\delta=k_d/k_m$. However, the resulting expressions become longer since the zeroth order terms cannot be grouped to retrieve the known isotropic results.

Finally we only have to proceed with the calculation of the anisotropic correction. As in the isotropic case, we choose the correlation direction along $z$ to simplify the integration. We consider two cases: correlation perpendicular to the sensor direction, and correlation parallel to it. For both cases, an arbitrary axis-of-symmetry of the medium is given by its spherical coordinates $(\theta_A,\varphi_A)$ or in Cartesian coordinates by $\mathbf{\hat{A}}=\alpha \mathbf{\hat{x}}+\beta\mathbf{\hat{y}}+\gamma\mathbf{\hat{z}}=\sin\theta_A\cos\varphi_A\mathbf{\hat{x}}+\sin\theta_A\sin\varphi_A\mathbf{\hat{y}}+\cos\theta_A\mathbf{\hat{z}}$.

\begin{enumerate}
	\item \textbf{Perpendicular correlation and sensor directions, i.e. $\boldsymbol{\mathbf{\theta_s}=\pi/2}$.} 
	Given that the correlation direction is along $z$, for $\mathbf{\theta_s}=\pi/2$, we choose the sensor axis to lie along $\mathbf{\hat{x}}$. Then, for an arbitrary axis-of-symmetry direction $\mathbf{\hat{A}}$, the integration of Eq. (\ref{eq:anisotropic_correction}) leads to
	\begin{equation}\begin{split}
	\delta B_{\perp}=-\frac{\delta_e}{4}\left\{\beta^2 \left[2j_2(k_o \Delta z)- j_0(k_o \Delta z) + \right.\right. \\ \left. \left.
	\cos(k_o \Delta z)\right]+\gamma^2j_2(k_o \Delta z)\right\}.
	\end{split}\end{equation}
	There is a harmonic term which does not decay with correlation distance, as would be expected. This is not a contradiction, but rather an artifact from the Taylor expansion: we are expanding the exponential and as correlation distance increases this first order approximation fails and more terms are needed. 
	
	Note that the component of the axis-of-symmetry along the sensor direction doesn't appear explicitly. This was expected since along the sensor axis the correction vanishes (extraordinary contribution becomes zero). Therefore varying $\alpha$ changes the magnitude of the correction, but doesn't alter its shape, which depends solely on the ratio between $\beta$ and $\gamma.$ The complete autocorrelation function becomes
	\begin{equation}\label{eq:CaseA}
	\begin{split}
	B_{V_sV_s}&=\frac{1}{2}\left( j_0(k_o\Delta z)-\frac{j_1(k_o\Delta z)}{k_o\Delta z}\right) \\
	-&\frac{\delta_e}{4}\bigg\{\cos^2\theta_A j_2(k_o \Delta z)- \\
	\sin^2\theta_A\sin^2\varphi_A \big[&2j_2(k_o \Delta z)- j_0(k_o \Delta z)+ \cos(k_o \Delta z)\big]\bigg\}.
	\end{split}\end{equation}
	
	Figure \ref{fig:GroupCaseA} shows the anisotropic result for different axis-of-symmetry orientations. When the axis-of-symmetry lies in the $yz$-plane ($\varphi_A=\pi/2$), i.e. $\mathbf{\hat{A}}=\sin\theta_A\mathbf{\hat{y}}+\cos\theta_A\mathbf{\hat{z}}$,  the sensor is perpendicular to both the axis-of-symmetry and correlation directions. This case corresponds to the maximum anisotropic contribution given any $\theta_A$. Even if departure of central lobes is not pronounced, their difference becomes significant after the first zero. On the other hand, whenever $\varphi_A=0$,  the axis-of-symmetry lies in the $xz$-plane as $\mathbf{\hat{A}}=\sin\theta_A\mathbf{\hat{x}}+\cos\theta_A\mathbf{\hat{z}}$, and the anisotropic contribution is the smallest (since the axis-of-symmetry projection on the sensor direction is the highest given a certain $\theta_A$). As seen in Fig. \ref{fig:GroupCaseA}, the autocorrelation function does not vary strongly for weak anisotropy in this configuration.
	
	\begin{figure}[!ht]
		\centering
		\includegraphics[width=0.85\linewidth]{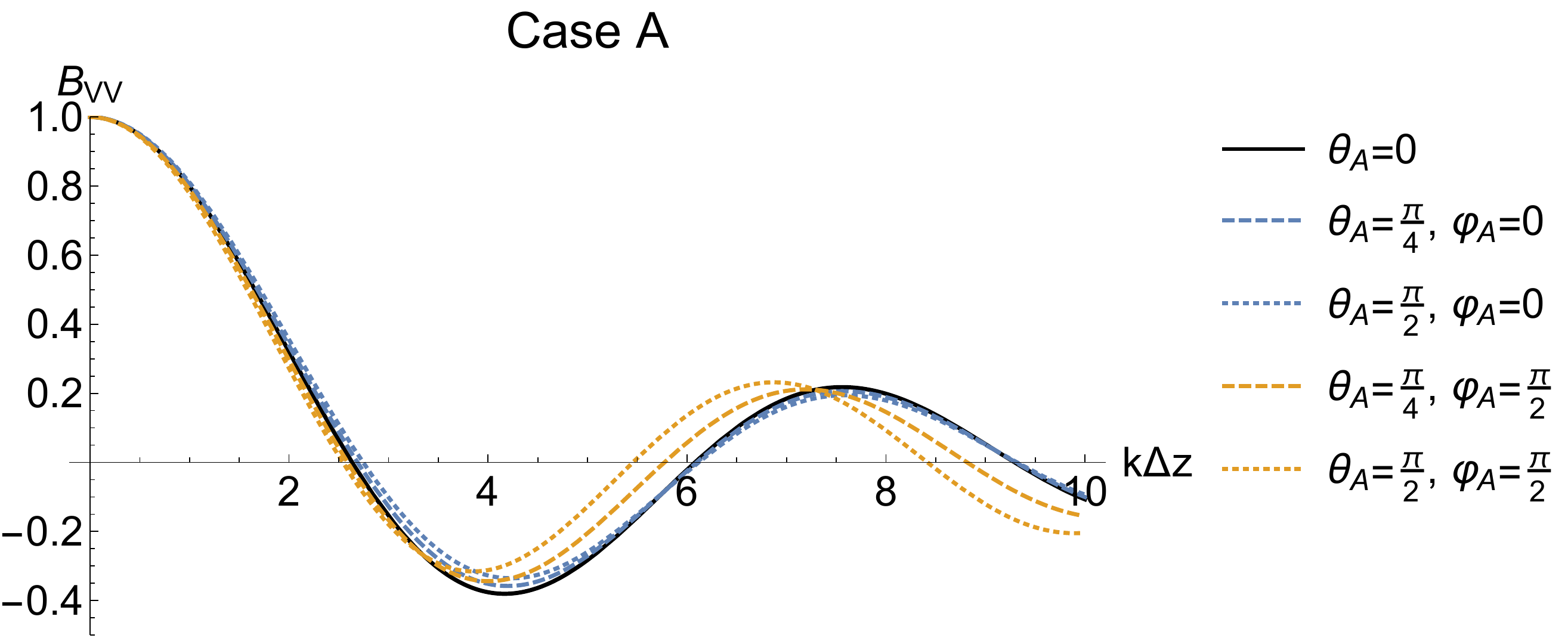}
		\caption{Autocorrelation function obtained with sensor perpendicular to correlation direction. Comparison of different axis-of-symmetry directions $(\theta_A,\varphi_A)$, and with a normalized anisotropy constant $\delta_e\approx0.23$. Scaling factor of $3$ is used for all the curves.}
		\label{fig:GroupCaseA}
	\end{figure}
	
	\item \textbf{Parallel correlation and sensor directions, i.e. $\boldsymbol{\mathbf{\theta_s}=0}$.}
	In this case, both the sensor and the correlation directions are along $z$; then $\mathbf{\hat{e}}_s=\mathbf{\hat{z}}$ and the integration of Eq. (\ref{eq:anisotropic_correction}) leads to
	\begin{equation}
	\delta B_{\parallel}=-\frac{\delta_e}{4}(1-\gamma^2)j_2(k_o\Delta 
	z).
	\end{equation}
	Hence the complete expression of the autocorrelation becomes
	\begin{equation}\label{eq:CaseB}
	B_{V_sV_s}=\frac{j_1(k_o\Delta 
		z)}{k_o\Delta 
		z}-\frac{\delta_e}{4}\sin^2(\theta_A)j_2(k_o\Delta 
	z),
	\end{equation}
	where, again, $\theta_A$ is the angle between the axis-of-symmetry and the correlation direction. Figure \ref{fig:GroupCaseB} shows the resulting autocorrelation for three different $\theta_A$ values, and $\delta_e\approx 0.23$. The central lobe width, given by the first zero position,  exhibits a small but noticeable change, greater than those in Figure \ref{fig:GroupCaseA}.
	
	\begin{figure}[!ht]
		\centering
		\includegraphics[width=0.8\linewidth]{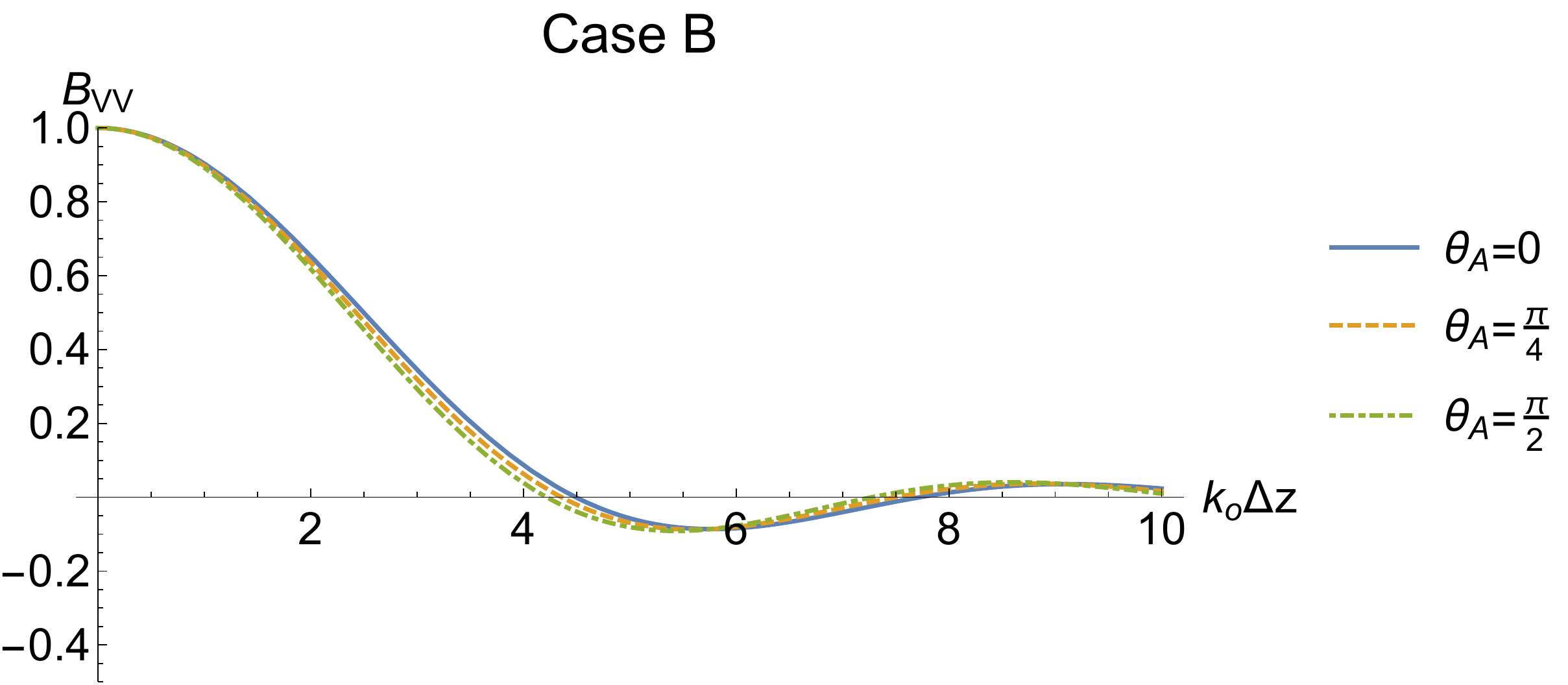}
		\caption{Autocorrelation function obtained for sensor parallel to both correlation directions. Comparison of three different $\theta_A$ when $\delta_e\approx0.23$. Scaling factor of 3 used for all the curves.}
		\label{fig:GroupCaseB}
	\end{figure}
	
\end{enumerate}

\subsection{Practical cases in USE and OCE}

In reverberant OCE \cite{Zvietcovich_2019}, the motion measurement  (sensor) direction is typically fixed along an axis, let us say the $x'$ axis, and 2D autocorrelations are taken along a plane perpendicular to it, the $y'z'$-plane. Then, Case A for $\varphi_A=\pi/2$, is of particular interest when the axis-of-symmetry of the material (e.g., orientation of fibers in muscle tissue) lies in the Y$'$Z$'$ plane at a certain $\theta_A$ angle. Here, $\theta_A$ is interpreted as the angle between the axis-of-symmetry and the correlation direction when the axis-of-symmetry is fixed to the $z'$ axis. Then, when $\theta_A=0$, the correlation direction corresponds to the $z'$ axis ($\Delta z'$), and when $\theta_A=\pi/2$, the correlation direction corresponds to the $y'$ axis ($\Delta y'$). 

\begin{figure}[!ht]
	\centering
	\includegraphics[width=\linewidth]{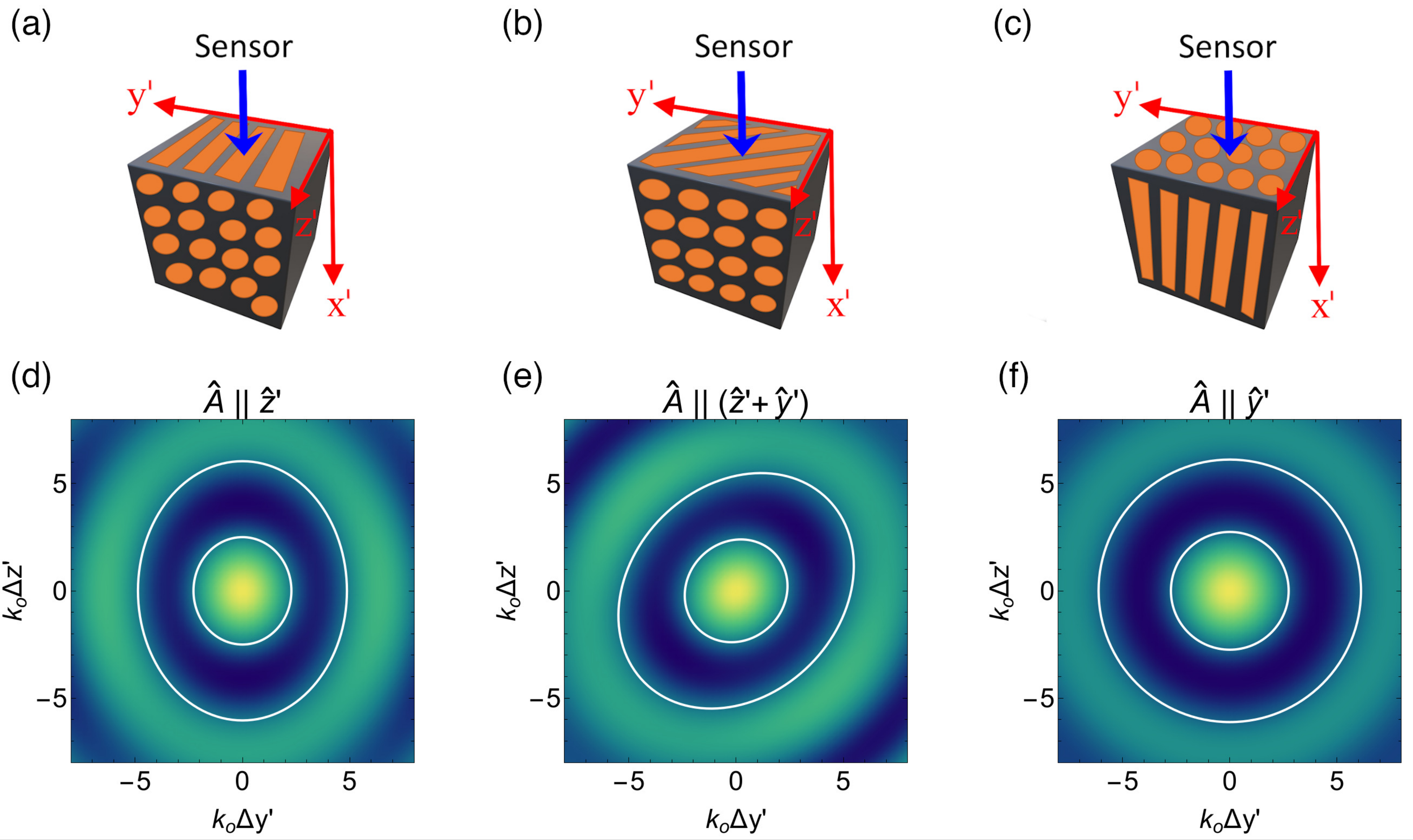}
	\caption{(a-c) Comparison between three different sample-sensor geometries having the sensor axis fixed along $x'$. Material's axis-of-symmetry: (a) parallel to $z'$; (b) at 45$^\circ$ from both $z'$ and $y'$; (c) along the sensor axis. (d-f) Resulting 2D autocorrelation maps in the $y'z$'-plane using the same scaling factor and $\delta_e\approx0.55$, corresponding to each geometry (a)-(c), i.e. axis-of-symmetry pointing at: (d) $z'$; (e) 45$^\circ$ from both $z'$ and $y'$; (f) parallel to $x'$.}
	\label{fig:2DCaseA2}
\end{figure}

In Figures \ref{fig:2DCaseA2}.a, and \ref{fig:2DCaseA2}.b, the full 2D autocorrelation maps are shown for two different axis-of-symmetry angles: parallel to $z'$ axis, and at 45$^\circ$ from both the $z'$ and the $x'$ axes. Then, by detecting the major and minor axes of the ellipses, not only the direction of fibers in muscle can be detected, but also their corresponding ordinary and extraordinary wave-numbers which are related to the shear modulus  parallel, and perpendicular to the fibers, respectively. When the axis-of-symmetry of the material is parallel to the sensor along $x'$ axis and 2D autocorrelations are taken along the $y'z'$-plane, Case A for $\varphi_A=0$ and $\theta_A=\pi/2$ is useful. As expected, in Figure \ref{fig:2DCaseA2}.c, the autocorrelation obtained is rotationally symmetric since plane $y'z'$, in this case, is the plane of isotropy in the transverse isotropic model of elasticity.

\begin{figure}[!ht]
	\centering
	\includegraphics[width=0.87\linewidth]{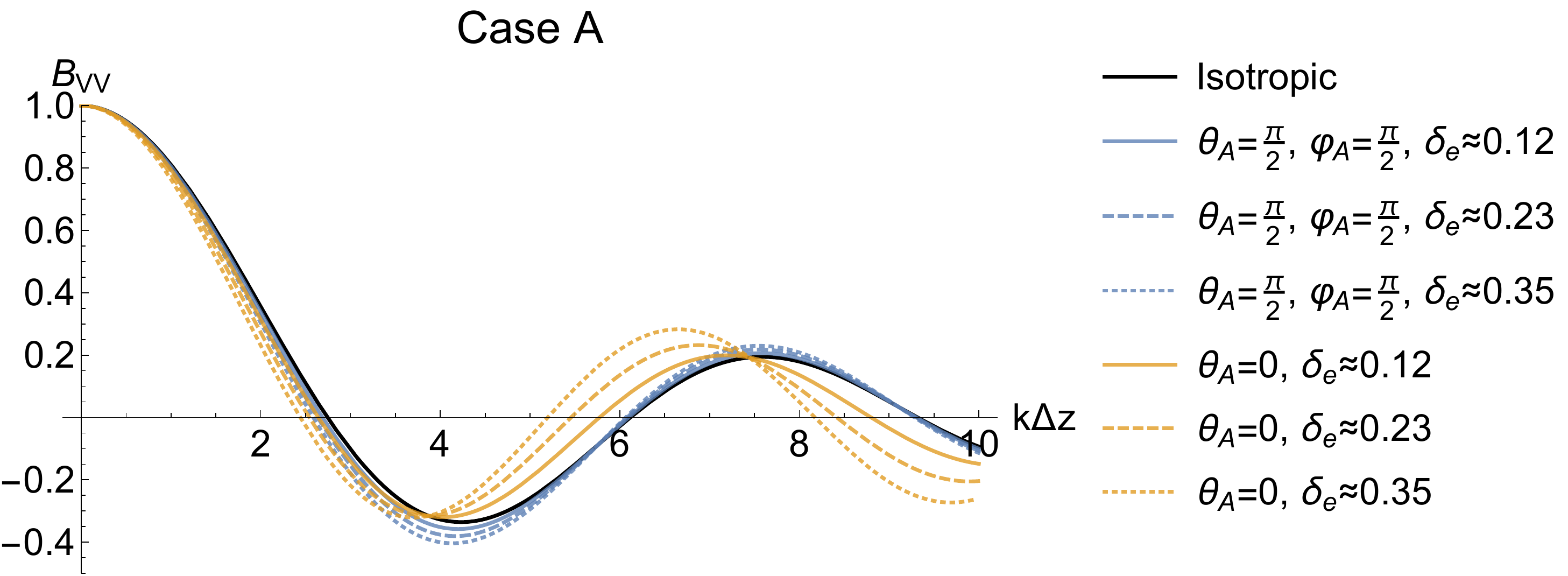}
	\caption{Comparison of Eq. (\ref{eq:CaseA}) ($\theta_s=\pi/2$, $\varphi_A=\pi/2$) for two canonical cases of axis-of-symmetry angles: $\theta_A=0$, and $\theta_A=\pi/2$ when the material has three different levels of anisotropy $\delta_e$. Curves are compared to the isotropic case using $k_o$.}
	\label{fig:CaseA90}
\end{figure}

We have derived the autocorrelation function for two cases, (A) correlation perpendicular to the sensor, and (B) correlation parallel to the sensor, given by Eqs. (\ref{eq:CaseA}) and (\ref{eq:CaseB}), respectively. Nevertheless, it is of interest to compare our results to earlier isotropic equations, since that has been the strategy used in previous work \cite{Ormachea_2018, Zvietcovich_2019}. Figure \ref{fig:CaseA90} shows the case for the sensor perpendicular to the axis-of-symmetry and correlation directions as in Figures \ref{fig:2DCaseA2}.a, and \ref{fig:2DCaseA2}.b. The comparison is made for the orthogonal cases $\theta_A=0$ (along fibers), and $\theta_A=\pi/2$ (perpendicular to fibers) for different values of anisotropy $\delta_e$ including the isotropic case using $k_o$. As shown, for a constant $k_o$, the larger the anisotropy, the larger the separation of the second lobe in the $\theta_A=0$ case with respect to the $\theta_A=\pi/2$ case. 

\begin{figure}[!ht]
	\centering
	\includegraphics[width=0.8\linewidth]{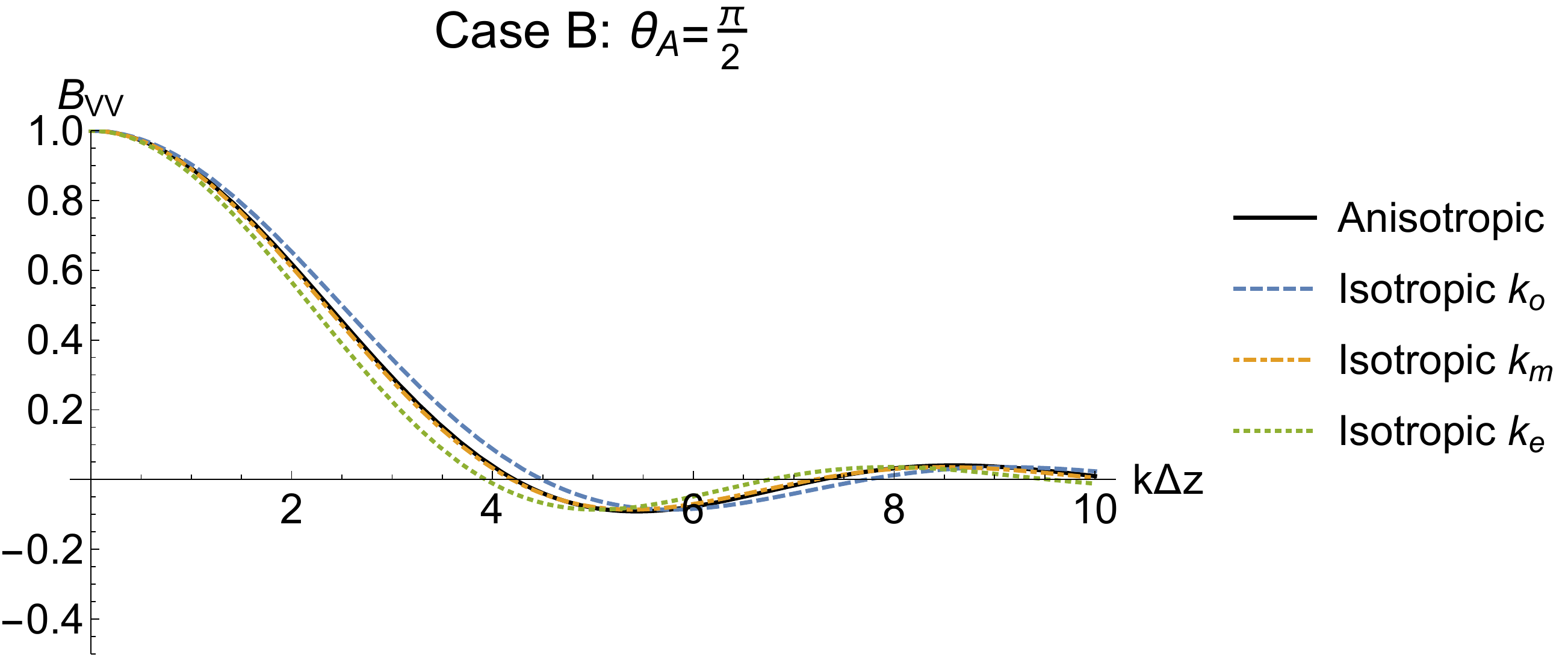}
	\caption{Comparison of three isotropic functions using $k_o$, $k_e$, and $k_m$, and the anisotropic expression up to first order. Here $\delta_e\approx0.23$ and $\theta_A=\pi/2$. Same scaling factor used for all the curves. Here the fitting by $k_m$ has a broader region of validity.}
	\label{fig:CaseB90}
\end{figure}

In reverberant USE \cite{Ormachea_2018}, when the motion measurement direction is typically located along the $x$ axis, due to USE capabilities in imaging larger depths, 2D autocorrelations are taken along the XY or XZ plane. Then, Case B is relevant. Figure \ref{fig:CaseB90} shows the comparison between the anisotropic result and three isotropic equations using $k_o$, $k_e$, and $k_m=(k_0 + k_e)/2$ for parallel sensor and correlation directions, and orthogonal axis-of-symmetry. Here, the isotropic equation using $k_m$ fits very well the central and side lobe. Therefore, $k_m$ in conjunction with the estimation of $k_o$ in Case A of Figure \ref{fig:CaseA90}, allows for the calculation of $k_e$.

Thus, as seen, the central region of the autocorrelation function can be fitted quite well using the isotropic expression. If birefringence is small, a more extended range is required to observe stronger differences, both in the zero positions and in the relative magnitude of side lobes. This explains why the isotropic theory was used successfully in the past for cornea\cite{Zvietcovich_2019}, although it is not isotropic \cite{PINSKY_2005,Singh_2016}. 

To fully implement the derived anisotropic autocorrelation, for example, one must first select the geometry of sensor-correlation (which in principle can always be chosen, although in practice may be restricted) and then fit the expression using 4 parameters: $k_o$, $\delta_e$, $\theta_A$, and $\varphi_A$. One measurement grants access to three different correlation directions (ideally many more since the correlation is done in 3D and interpolation could be employed to obtain profiles at other angles) which can be used together to determine the anisotropy of the system as well as the axis orientation without any \emph{a priori} assumption of the axis-of-symmetry direction.

\section{Numerical simulations}

\subsection{Simulation setup}
\begin{figure}[!ht]
	\centering
	\includegraphics[width=1\linewidth]{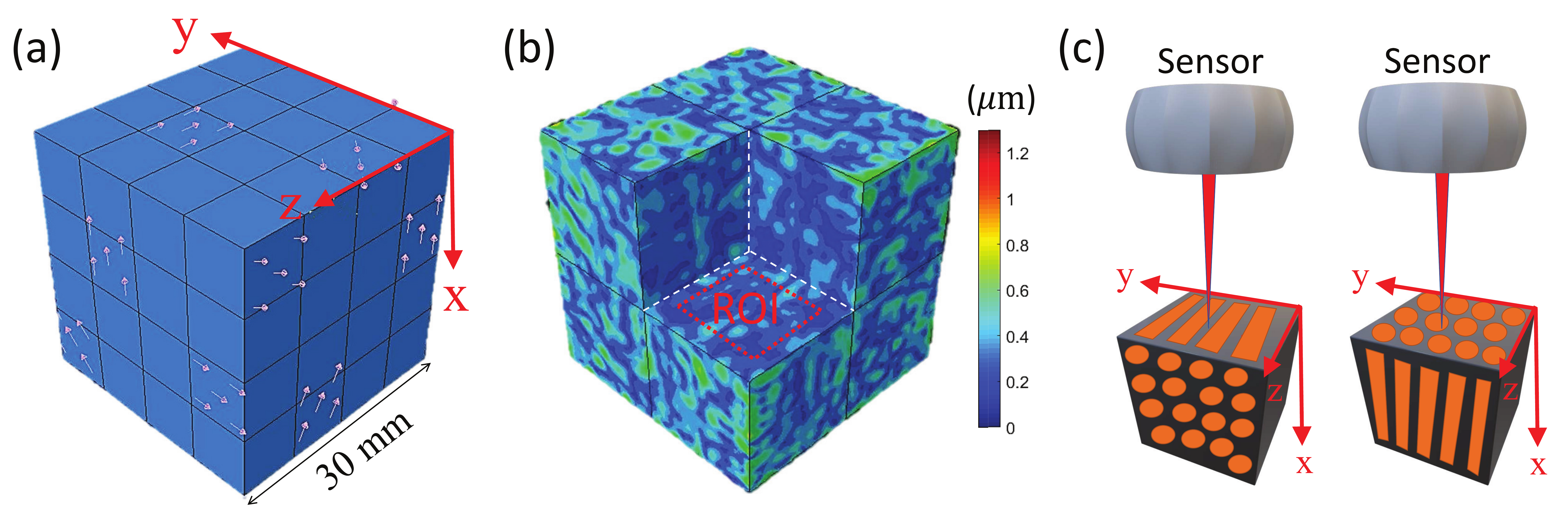}
	\caption{Numerical simulation of a reverberant shear wave field in anisotropic media. (a) Dimensions and boundary conditions of a 3D solid subjected to multiple shear sources vibrating at 2700 Hz. (b) Displacement magnitude field (color bar in $\mu$m) measured along the $x$ axis after simulation. (c) Cases of axis-of-symmetry orientation of the material along the $z$ axis (left), and $x$ axis (right).}
	\label{fig:CaseAbaqus}
\end{figure}

Numerical simulations of a reverberant shear wave field produced by multiple shear-displacement contacts applied to the surface of a 3D solid volume were conducted using finite elements in Abaqus/CAE version 6.14-1 (Dassault Systems, Velizy-Villacoublay, France). The 3D solid of 30 x 30 x 30 mm is subjected to spatially-uniform (square shape) and temporal-harmonic (2700 Hz) displacement field at different surface locations as shown in Figure \ref{fig:CaseAbaqus}a. Zero displacement and rotation were applied at the base of the cube. The solid was meshed with an approximate grid size of 0.1 mm and using linear hexahedral dominant elements (C3D8R). The type of simulation was selected to be steady-state dynamic direct. After the simulation, a 3D complex-valued displacement field along the $x$ axis (sensor axis) is extracted as shown in Figure \ref{fig:CaseAbaqus}b. Finally, the complex autocorrelation is evaluated in regions of interest (ROI) of 18 mm x 18 mm along the YZ plane throughout the 3D displacement volume.

\subsection{Material properties}
The solid material is represented using a linear and transverse isotropic model of elasticity with a density of $\rho$ = 1000 kg/m$^{3}$ and parameters defined in Table \ref{tab:Table1}. In this model, the material properties are symmetric within the plane-of-isotropy ($p$), which is perpendicular to the axis-of-symmetry ($t$) direction (also called direction of fibers in muscle). The compliance tensor of a transverse isotropic material can be represented with the following 7 parameters: $E_p$, and $E_t$, corresponding to the Young?s moduli in the plane-of-isotropy and along the axis-of-symmetry, respectively; $G_p$, and $G_t$, corresponding to shear moduli in the plane-of-isotropy, and in a transverse plane parallel to the axis-of-symmetry, respectively; and $\nu_p$, $\nu_{pt}$ (and $\nu_{tp}$), corresponding the the Poisson's ratios in the plane-of-isotropy, and two transverse planes parallel to the axis-of-symmetry, respectively. Finally, these variables can be reduced to 3 independent parameters if the material is considered incompressible (such as soft tissues) \cite{Itskov_2002}.

\begin{table}[h!]
	\caption{Material parameters using the transverse isotropic model defined in Abaqus/CAE version 6.14-1. Elastography parameters are also calculated for further comparison.}
	\label{tab:Table1}
	\centering\includegraphics[width=1\linewidth]{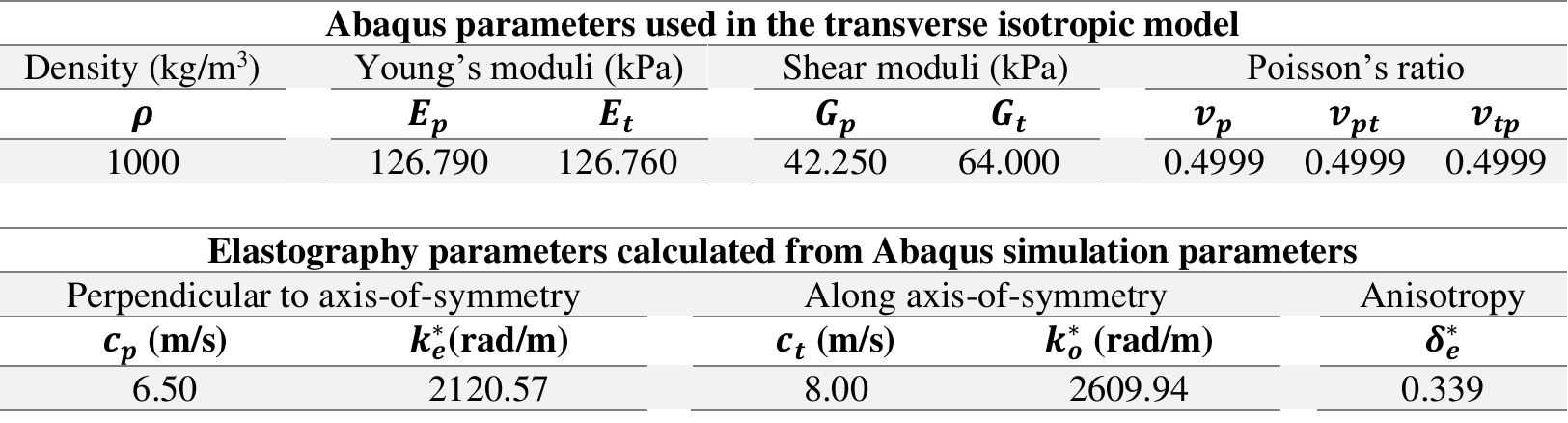}
\end{table}

In dynamic elastography, we are interested in the propagation of shear waves, leaving $G_p$ and $G_t$ as the most important parameters since they can be related to shear wave speeds $c_p$ and $c_t$, using $c_p=\sqrt{G_p/\rho}$ and $c_t=\sqrt{G_t/\rho}$, respectively \cite{Royer_2011}. On the other hand, in reverberant elastography \cite{Ormachea_2018, Zvietcovich_2019}, for a vibration frequency $f$, wave-numbers are typically estimated. Then, $G_p$ and $G_t$ can be related to the extraordinary $k_e$ and ordinary $k_o$ wave-numbers using $k_e=2\pi f/\sqrt{G_p/\rho}$ and $k_o=2\pi f/\sqrt{G_t/\rho}$, respectively. Calculations of these wave-numbers for $f$ = 2700 Hz, and shear wave speeds, based on the simulation parameters, are also reported in Table \ref{tab:Table1}. In reverberant OCE \cite{Zvietcovich_2019}, the sensor is usually fixed in one axis, and autocorrelations are taken along a plane perpendicular to the sensor. Then, we define the $x$ axis as the sensor direction, and the YZ plane as the autocorrelation plane. Two cases are explored: (Case 1) when the axis-of-symmetry is oriented along the $z$ axis (Figure \ref{fig:CaseAbaqus}c-left), and (Case 2) when the axis-of-symmetry is oriented along the $x$ axis (Figure \ref{fig:CaseAbaqus}c-right in which the autocorrelation plane is also the plane-of-symmetry).

\subsection{Results and discussion}

\begin{figure}[!ht]
	\centering
	\includegraphics[width=1\linewidth]{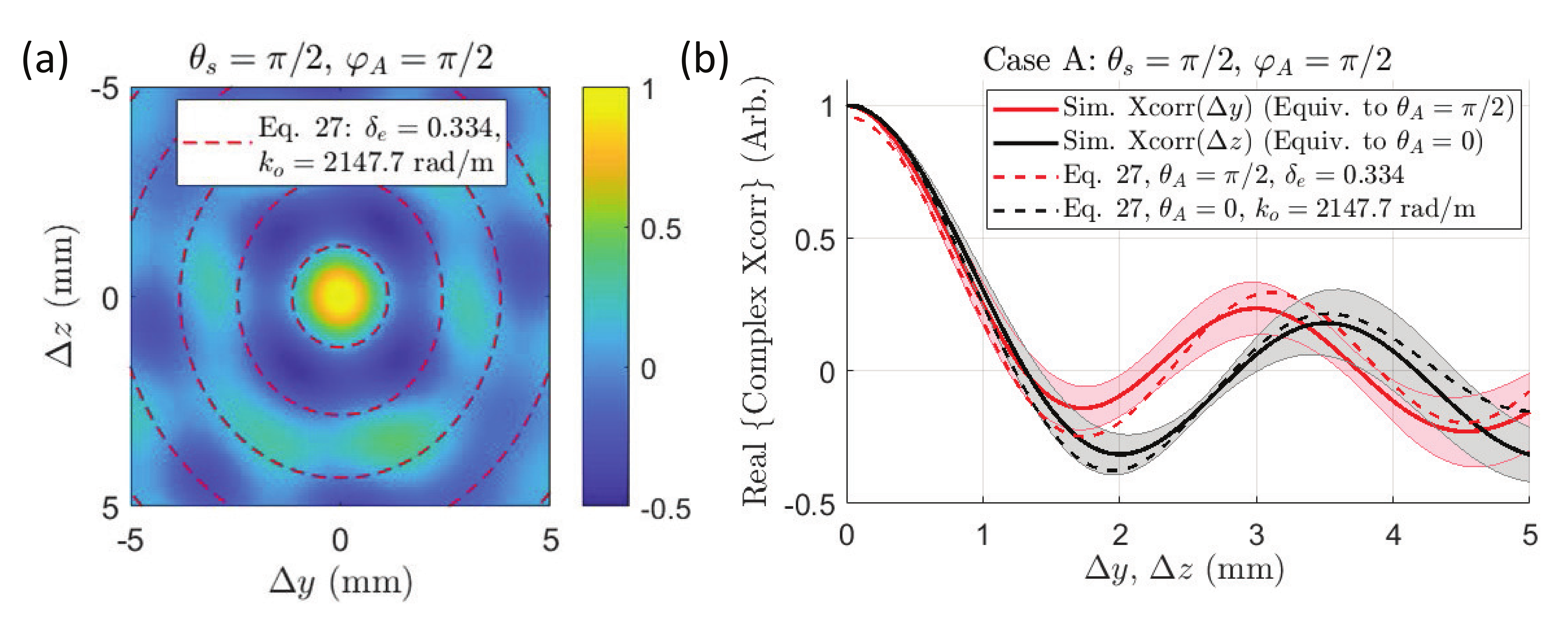}
	\caption{Fitting of Eq. (\ref{eq:CaseA}) with simulation results in Case 1. (a)  2D average autocorrelation along the YZ plane, obtained from the simulated 3D displacement volume, is fitted to Eq. (\ref{eq:CaseA}) for $\theta_s=\pi/2$ and $\varphi_A=\pi/2$ (discontinuous red line representing the zeros of Eq. (\ref{eq:CaseA})). Colorbar represents normalized autocorrelation in arbitrary units. (b) Major and minor axes of the ellipse corresponding to $\Delta z$ and $\Delta y$ autocorrelation axes, respectively, are compared against simulation results.  Fitting parameters $k_o$ = 2147.7 rad/m and $\delta_e$ = 0.334 were estimated providing a close match to the ground truth.}
	\label{fig:SimFit_A}
\end{figure}

In Case 1, the average 2D autocorrelation calculated from ROIs along the YZ plane of the 3D displacement volume is fitted to Eq. (\ref{eq:CaseA}) ($\theta_s=\pi/2$) when $\varphi_A=\pi/2$ (Figure \ref{fig:SimFit_A}a). Here, $\theta_A$ is interpreted as the angle between the axis-of-symmetry and the correlation direction when the axis-of-symmetry is fixed to the $z$ axis. Then, when $\theta_A=0$, the correlation direction corresponds to the $z$ axis ($\Delta z$), and when $\theta_A=\pi/2$, the correlation direction corresponds to the $y$ axis ($\Delta y$). An elliptical shape in the plot is clearly observed in Figure \ref{fig:SimFit_A}a indicating that the anisotropic properties of the material are different parallel ($\Delta z$) and perpendicular ($\Delta y$) to the axis-of-symmetry. The major and minor axes of the ellipse corresponding to the $\Delta z$ and $\Delta y$ autocorrelation axes, respectively, are shown with Eq. (\ref{eq:CaseA}) ($\theta_s=\pi/2$) curve fittings in Figure \ref{fig:SimFit_A}b. Fitting parameters $k_o$ and $\delta_e$ are shown and compared against simulation ground truth parameters in Table \ref{tab:Table2}.

\begin{table}[h!]
	\caption{Estimated ordinary and extraordinary wave-numbers based on the fitting parameters $k_o$, and $\delta_e$ in Case 1 and 2. Average parameters are compared against ground truth parameters set in the simulation (Table \ref{tab:Table1}).}
	\label{tab:Table2}
	\centering\includegraphics[width=1\linewidth]{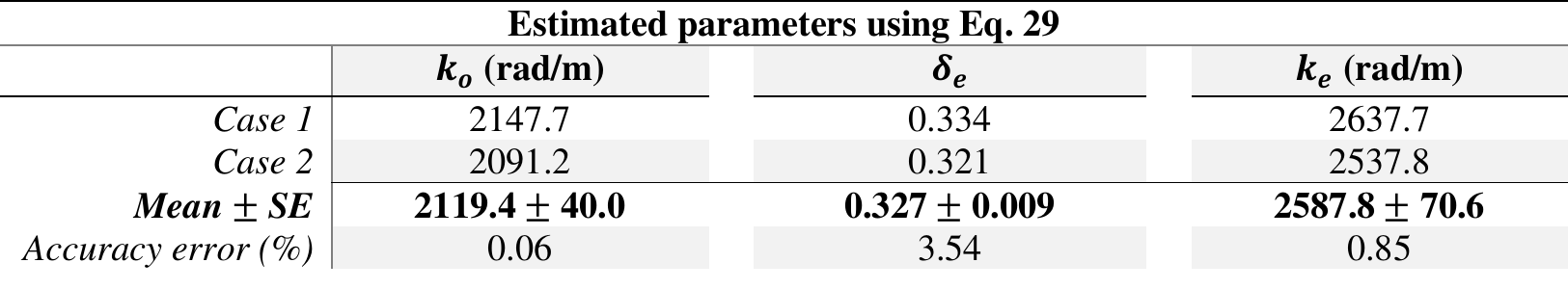}
\end{table}

\begin{figure}[!ht]
	\centering
	\includegraphics[width=1\linewidth]{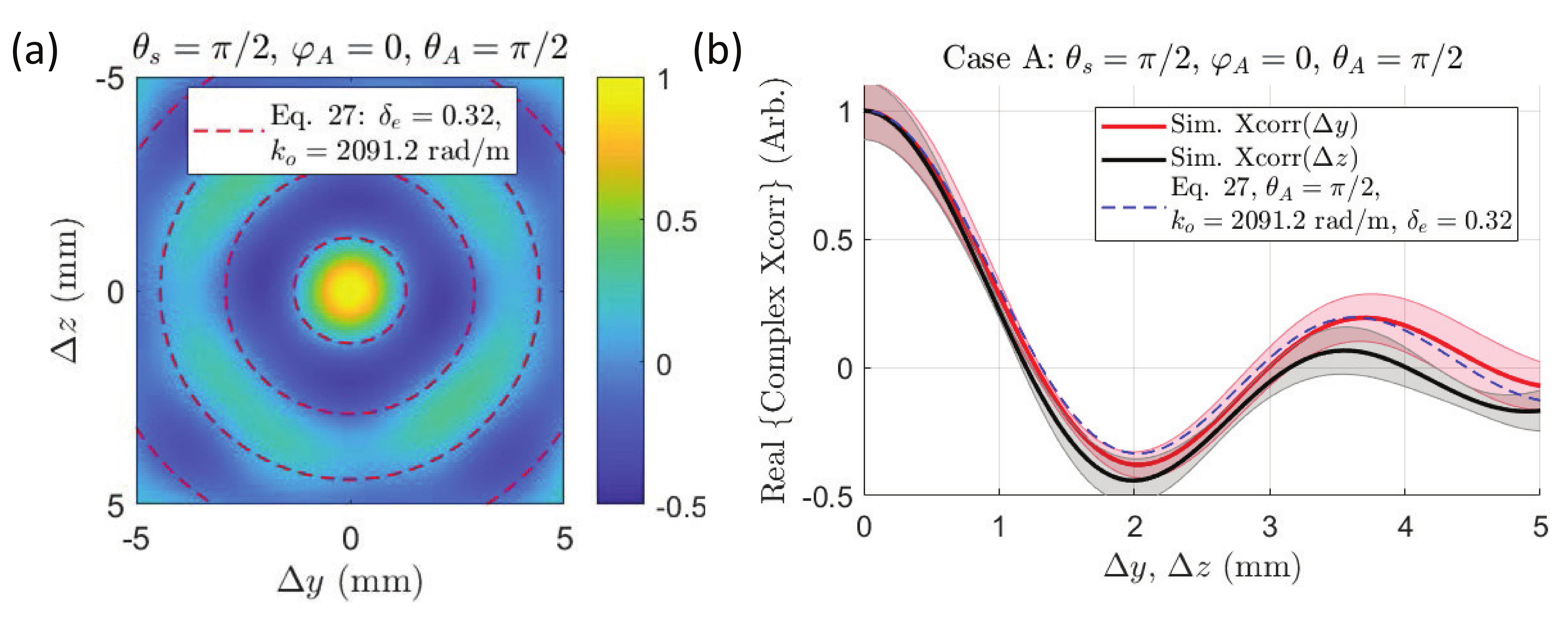}
	\caption{Fitting of Eq. (\ref{eq:CaseA}) with simulation results in Case 2. (a)  2D average autocorrelation along the YZ plane, obtained from the simulated 3D displacement volume, is fitted to Eq. (\ref{eq:CaseA}) for $\theta_s=\pi/2$, $\varphi_A=0$, and $\theta_A=\pi/2$ (discontinuous red line representing the zeros of Eq. (\ref{eq:CaseA})). Colorbar represents normalized autocorrelation in arbitrary units. (b) Autocorrelation axes $\Delta z$ and $\Delta y$ are compared against simulation results.  Fitting parameters $k_o$ = 2091.2 rad/m and $\delta_e$ = 0.320 were estimated, providing a good assessment of the material properties used in the simulation.}
	\label{fig:SimFit_A2}
\end{figure}

Similarly, in Case 2, the average 2D autocorrelation is taken along the YZ plane when the axis-of-symmetry is oriented along the $x$ axis and fitted to Eq. (\ref{eq:CaseA}) ($\theta_s=\pi/2$) when $\varphi_A=0$ and $\theta_A=\pi/2$ (Figure \ref{fig:SimFit_A2}a). Here, the interpretation of $\theta_A$ is the same as in Section 3.3. As expected, the plot shape is circular and symmetric as Eq. (\ref{eq:CaseA}) in this case is the same for any correlation direction perpendicular to the sensor and axis-of-symmetry directions. Autocorrelation axes along $\Delta z$ and $\Delta y$ are shown with Eq. (\ref{eq:CaseA}) ($\theta_s=\pi/2$, $\varphi_A=0$, and $\theta_A=\pi/2$) curve fittings in Figure \ref{fig:SimFit_A2}b. Fitting parameters $k_o$ and $\delta_e$ are shown and compared against simulation ground truth parameters in Table \ref{tab:Table2}. 

Estimations of $k_o$ and $\delta_e$ are used in Eq. (\ref{eq:Delta_e}) for the calculation of $k_e$ in each case as reported in Table \ref{tab:Table2}. Average estimations are compared against ground truth parameters set in the simulation (Table \ref{tab:Table1}). We found a maximum accuracy error of 3.54\% and a minimum of 0.06\%, validating the effectiveness of the anisotropic derivation in reverberant shear wave fields. This has important implications in the elastography of transverse isotropic elastic tissues: (1) the axis-of-symmetry of tissues (for example the fiber direction in muscle) can be estimated by finding the major axis of the elliptical plot of Eq. (\ref{eq:CaseA}) in Case 1; (2) the complete characterization of shear moduli in every direction ($G_p$, and $G_t$) can estimated based on $k_o$ and $\delta_e$ provided by Eq. (\ref{eq:CaseA}) in Cases 1 and 2; and (3) more complex situations in which the axis-of-symmetry of the tissue is not parallel to one of the axes can be fully characterized by building libraries of cases using Equations (\ref{eq:CaseA}) and (\ref{eq:CaseB}) and machine learning tools.

\section{Reverberant OCE experiments}

\subsection{Sample preparation}
\begin{figure}[!ht]
	\centering
	\includegraphics[width=1\linewidth]{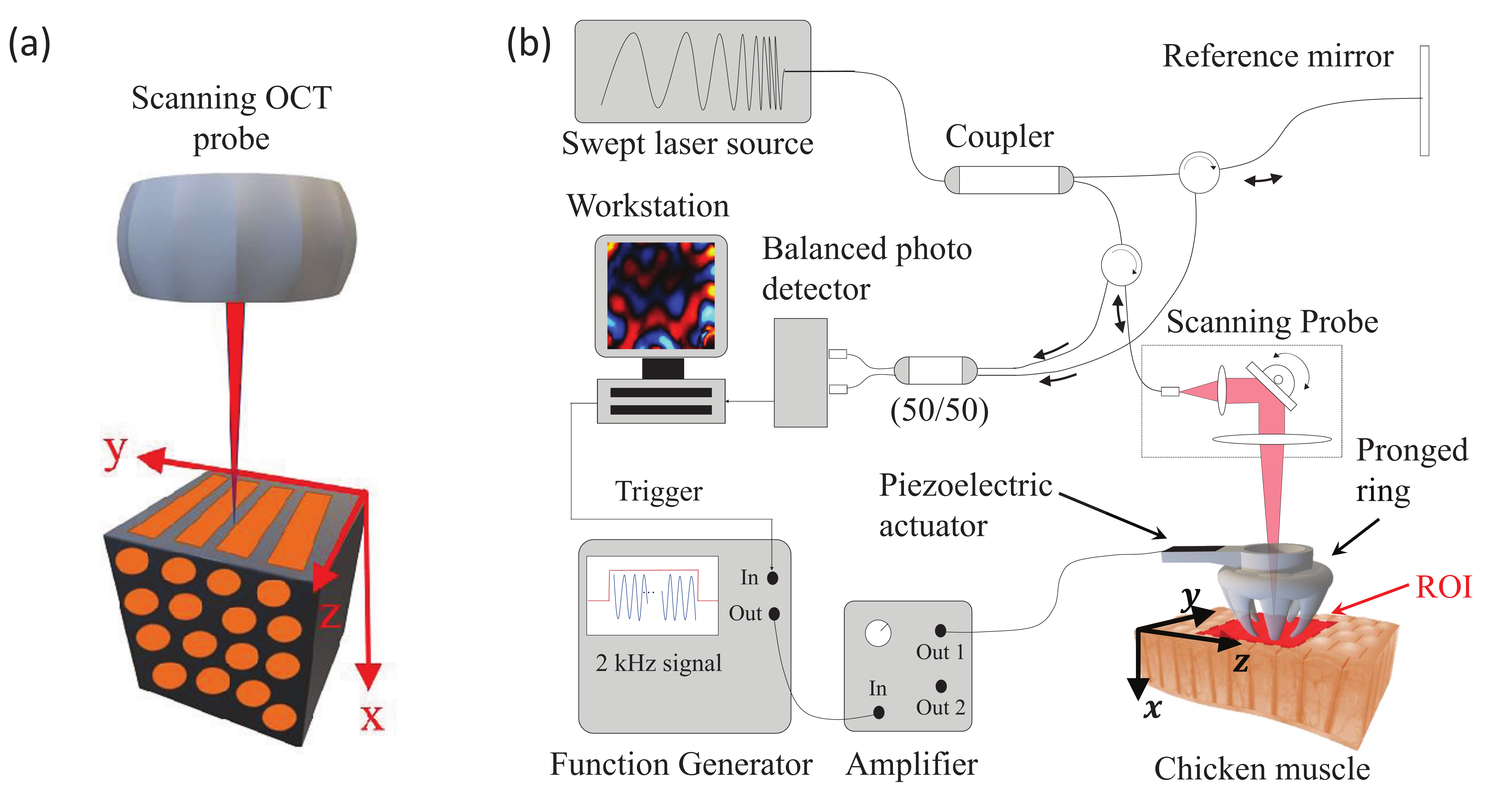}
	\caption{Experimental opto-mechanical setup for the generation and measurements of reverberant shear wave fields in chicken muscle tissue. (a) Orientation of the chicken muscle sample with respect to the OCT scanning probe. Average orientation of fibers was aligned along the $z$ axis, while the motion measurement (sensor) was oriented along the $x$ axis (depth). (b) Phase-sensitive OCT system based on a swept source laser. A 2 kHz mechanical excitation was generated in the sample using a 3D printed pronged ring allowing for motion measurement along the $yz$-plane within the ROI (9 mm x 9 mm). }
	\label{fig:ExpSetup}
\end{figure}

Using a surgical scalpel, three (n = 3) cubical samples (2 x 2 x 2 cm) were dissected from a fresh roaster chicken tibialis anterior muscle. Each cubical sectioning was conducted so that the fiber orientation of the muscle is parallel to one of the axes of the cube. The epithelium was removed from all sides of the cubic sample since OCE measurements are usually constrained to the surface of the sample. During experiments, the side of the cubical sample containing all fibers oriented to one of the axes of the cube was measured (Figure \ref{fig:ExpSetup}a). The muscle was not subjected to any external force in order to prevent a passive muscle resistance effect.

\begin{figure}[!ht]
	\centering
	\includegraphics[width=\linewidth]{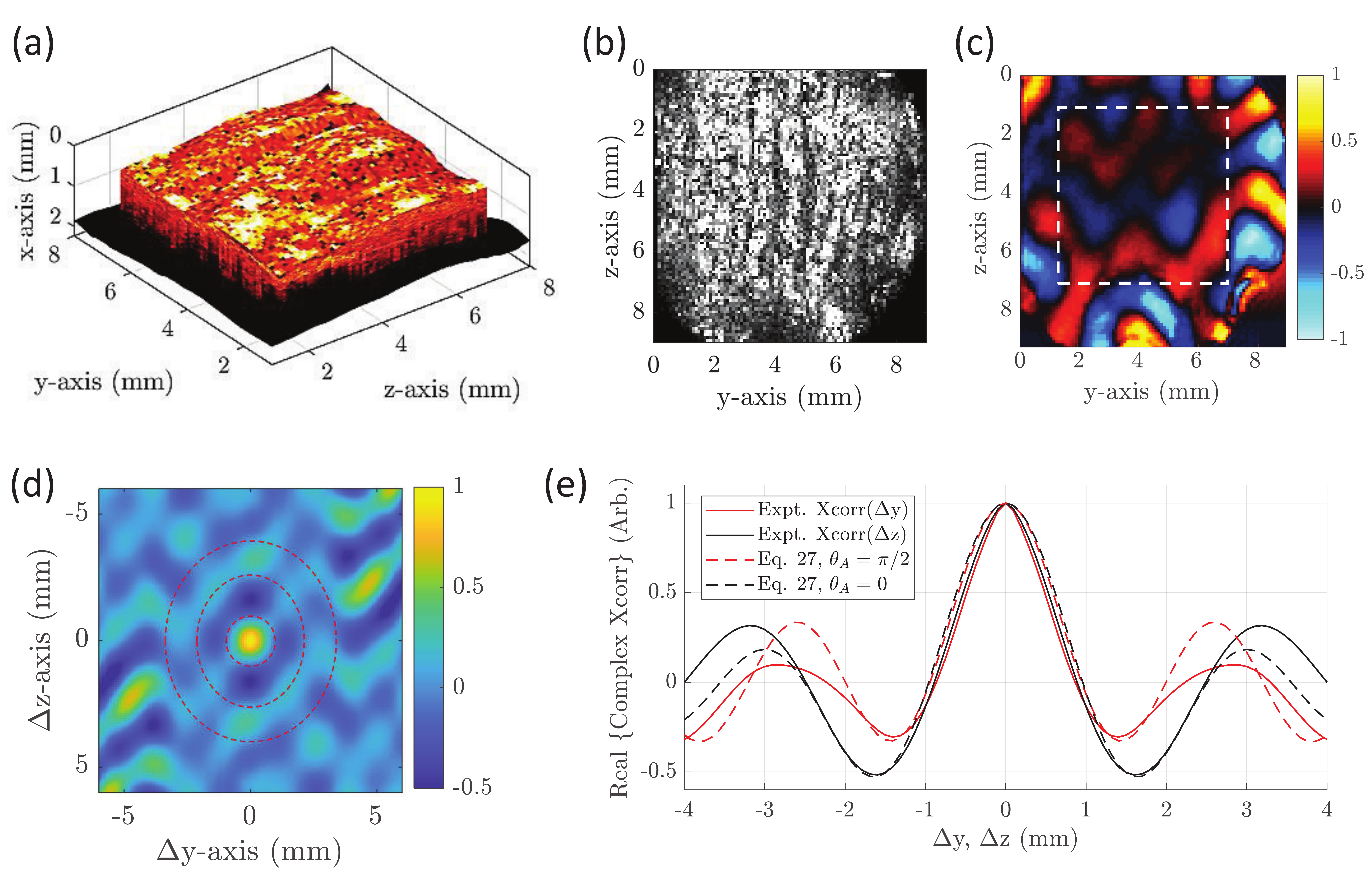}
	\caption{Experimental reverberant OCE results in chicken muscle. (a) 3D structural OCT volume of one of the muscle samples. (b) Structural \emph{en face} OCT image of the muscle along the $yz$-plane. Color map represents normalized intensity. (c) Motion snapshot of a 2 kHz reverberant field measured at the surface of the muscle sample at $t_0$ = 2ms instant. Color bar represents normalized particle velocity in arbitrary units. (d) 2D autocorrelation of the reverberant field extracted from a 6 mm x 6 mm region (white discontinuous line) in (c). Color bar represents the normalized real part of the complex autocorrelation in arbitrary units. Discontinuous red line represents the zeros of Eq.(\ref{eq:CaseA}) for $\theta_s=\pi/2$ and $\varphi_A=\pi/2$. (e) Major ($\Delta z$) and minor ($\Delta y$) autocorrelation axis of the ellipse in (d) fitted to Eq. (\ref{eq:CaseA}) for cases $\theta_A=0$ and $\theta_A=\pi/2$, respectively. Fitting parameters $k_o$ = 2512.3 rad/m and $\delta_e$ = 0.42 were estimated for muscle sample 1. Fitting quality: $r^2$ = 0.962.}
	\label{fig:ExpResults}
\end{figure} 

\subsection{Experimental setup and processing scheme}
The experimental setup consists of a phase-sensitive optical coherence tomography (PhS-OCT) system implemented with a swept source laser (HSL-2100-WR, Santec, Aichi, Japan) of a center wavelength of 1318 nm and a bandwidth of 125 nm (Figure \ref{fig:ExpSetup}b). The frequency sweep rate of the light source was 20 kHz, and the optical resolution was measured to be 30 $\mu$m laterally, and 10 $\mu$ m axially. The system was used to acquire 3D motion frames of the chicken samples within a ROI of 9 x 9 mm in the YZ-plane. The mechanical excitation system begins with a function generator (AFG320, Tektronix, Beaverton, OR, USA) output signal connected to an ultra-low noise power amplifier (PDu150, PiezoDrive, Callaghan, NSW, Australia) feeding a piezoelectric bender poled in a parallel configuration of 10 x 45 mm surface dimensions (BA4510, PiezoDrive, Callaghan, NSW, Australia). A 3D printed pronged ring containing eight vertical equidistant and circular distributed rods is attached to one of the ends of the piezoelectric bender (Figure \ref{fig:ExpSetup}b). The rods are lightly touching the sample surface in a concentric configuration and produce a reverberant field when the piezoelectric bender is excited at 2 kHz. The ring shape allows the imaging of the cornea using the OCT system, while the rods introduce mechanical excitation. Reverberant particle velocity (motion) fields along the $x$ axis (sensor axis) were analyzed in the $yz$-plane in order to calculate complex 2D autocorrelations for further fitting with Eq. (\ref{eq:CaseA}). Anisotropic properties of the $n = 3$ chicken muscle samples were characterized by estimating parameters $k_o$ and $\delta_e$ as conducted in Section 4.3 for the simulated case.

\subsection{Results and discussion}
Figure \ref{fig:ExpResults}a shows the 3D structural OCT volume of one of the chicken samples. The average direction of the muscle fibers is aligned toward the $z$ axis as shown in the \emph{en face} structural image of Figure \ref{fig:ExpResults}b taken along the $yz$-plane. A motion snapshot (normalized particle velocity in arbitrary units) of the 2 kHz reverberant field produced in the chicken sample is shown in Figure \ref{fig:ExpResults}c. Here, a 6 x 6 mm region was selected for the calculation of the 2D autocorrelation (normalized units) and fitted to Eq. (\ref{eq:CaseA}) ($\theta_s=\pi/2$) when $\varphi_A=\pi/2$ (Figure \ref{fig:ExpResults}d). An elliptical shape in Figure \ref{fig:ExpResults}d highlights the anisotropic properties of muscle tissue when comparing autocorrelation plots parallel ($\Delta z$) and perpendicular ($\Delta y$) to the $z$ axis. The major and minor axes of the ellipse corresponding to $\Delta z$ and $\Delta y$ autocorrelation axes, respectively, are fitted to Eq. (\ref{eq:CaseA}) in Figure \ref{fig:ExpResults}e. Fitting parameters $k_o$ and $\delta_e$ are estimated and shown for all samples in Table \ref{tab:Table3}.

\begin{table}[h!]
	\caption{Estimated shear moduli in the plane-of-isotropy (XY-plane) $G_p$ and in the transverse plane parallel to the axis-of-symmetry ($z$ axis) $G_t$, based on the fitting parameters $k_o$, and $\delta_e$ in $n = 3$ chicken muscle samples. Shear wave speed was also calculated along the same directions for further comparison. SE: standard error.}
	\label{tab:Table3}
	\centering\includegraphics[width=1\linewidth]{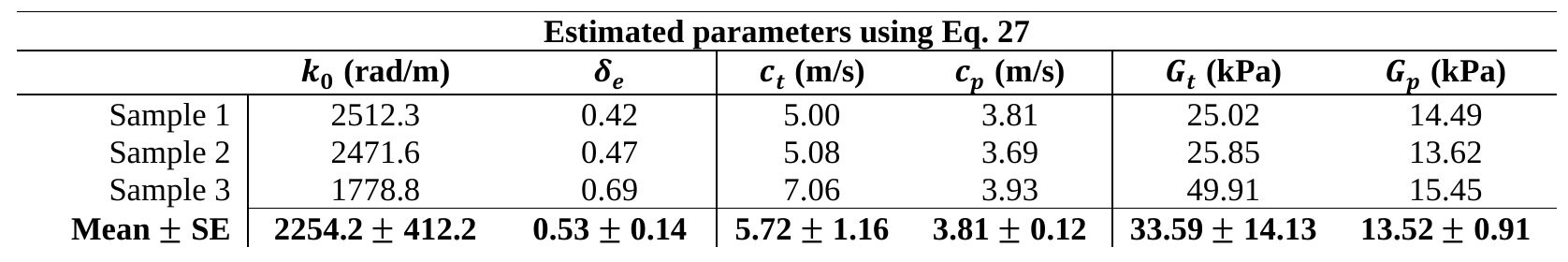}
\end{table}

As explained in Section 4.2, for a transverse isotropic medium, the shear moduli in the plane-of-isotropy $G_p$ and in the transverse plane parallel to the axis-of-symmetry (direction of the fibers) $G_t$ can be calculated from shear speed $c_p$ and $G_t$, respectively, using $k_o$ and $\delta_e$ parameters. Table \ref{tab:Table3} shows $c_p$, $c_t$, $G_p$, and $G_t$ for all chicken samples, indicating a marked anisotropy in agreement with other studies \cite{Zvietcovich_2020,Koo_2013}. The fitting quality of Eq. (\ref{eq:CaseA}) to autocorrelation plots tends to degrade as sample points are further away from the center of the autocorrelation (Figure \ref{fig:ExpResults}e). This is explained as Eq. (\ref{eq:CaseA}) comes from a theoretical formulation of autocorrelation considering an infinite space field (numerous spatial waves within a region). In practice, due to the attenuation of waves in tissues, a limited number of cycles can be captured within a ROI as shown in Figure \ref{fig:ExpResults}c, constraining the effectiveness of the fitting of Eq. (\ref{eq:CaseA}) to the center of the autocorrelation map in Figure \ref{fig:ExpResults}d. Finally, this study demonstrates that reverberant elastography can be used in practical cases for the characterization of transverse isotropic tissues such as muscle. Future work will focus on extending this method to other anisotropic tissues such as cornea and brain.

\section{Conclusion}
The major concepts from electromagnetic fields in anisotropic media are reviewed and found to be helpful in deriving closed-form solutions to the problem of reverberant elastography in anisotropic media. We found Equations (\ref{eq:CaseA}) and (\ref{eq:CaseB}) describing the complex autocorrelation of reverberant fields in materials exhibiting a transverse isotropic model of elasticity for variable directions of: (1) the material's axis-of-symmetry, (2) the direction of motion measurement (sensor), and (3) complex autocorrelation. Results were validated with numerical simulations using finite elements achieving accuracy within 4\%. Moreover, Equation (\ref{eq:CaseA}) was used for the anisotropic characterization of chicken tibialis anterior muscle in OCE experiments, demonstrating its use in the non-destructive elastography of tissues. Finally, we developed a general solution for the isotropic model in Eq. (\ref{eq:isotropic_result}) consistent with previous reported results for particular configurations. Limitations of this work include the assumption of small anisotropic ratios and the consequent simplification of terms within the complex autocorrelation function. Future work will focus on the application of this approach to the elastography of other well know anisotropic tissues such as cornea and brain.

\section*{Acknowledgment}
The authors would like thank Prof. Miguel Alonso for his perspective. L. A. Alem{\'a}n-Casta{\~n}eda is supported by CONACyT Doctoral Fellowship, and F. Zvietcovich was supported by the Fondo para la
Innovacion, la Ciencia y la Tecnologia FINCyT--Peru (097-FINCyT-BDE-2014).

\bibliographystyle{ieeetr}
\bibliography{anisotropy_optics,elastography,anisotropy_biological}

\begin{thebibliography}{10}

\bibitem{Parker_2010}
K.~Parker, M.~Doyley, and D.~J. Rubens, ``Imaging the elastic properties of
  tissue: the 20 year perspective,'' {\em Physics in Medicine and Biology},
  vol.~56, no.~1, pp.~R1--R29, 2010.

\bibitem{Feng_2013}
Y.~Feng, R.~J. Okamoto, R.~Namani, G.~M. Genin, and P.~V. Bayly, ``Measurements
  of mechanical anisotropy in brain tissue and implications for transversely
  isotropic material models of white matter,'' {\em J Mech Behav Biomed Mater},
  vol.~23, pp.~117--32, 2013.

\bibitem{LEVINSON_1987}
S.~F. Levinson, ``Ultrasound propagation in anisotropic soft tissues: The
  application of linear elastic theory,'' {\em Journal of Biomechanics},
  vol.~20, no.~3, pp.~251 -- 260, 1987.

\bibitem{Yariv:book}
A.~Yariv and P.~Yeh, {\em Optical Waves in Crystals}.
\newblock John Wiley $\&$ Sons, 1983.
\newblock {Chapter} 4.

\bibitem{Aleman2016}
L.~A. Alem\'an-Casta\~neda and M.~Rosete-Aguilar, ``Deviation from orthogonal
  polarization for ordinary and extraordinary rays in uniaxial crystals,'' {\em
  Journal of the Optical Society of America A}, vol.~33, no.~4, pp.~677--682,
  2016.

\bibitem{Gennisson_2010}
J.-L. Gennisson, T.~Deffieux, E.~Macé, G.~Montaldo, M.~Fink, and M.~Tanter,
  ``Viscoelastic and anisotropic mechanical properties of in vivo muscle tissue
  assessed by supersonic shear imaging,'' {\em Ultrasound in Medicine \&
  Biology}, vol.~36, no.~5, pp.~789--801, 2010.

\bibitem{Royer_2011}
D.~Royer, J.-L. Gennisson, T.~Deffieux, and M.~Tanter, ``On the elasticity of
  transverse isotropic soft tissues (l),'' {\em The Journal of the Acoustical
  Society of America}, vol.~129, no.~5, pp.~2757--2760, 2011.

\bibitem{Wang_2013}
M.~{Wang}, B.~{Byram}, M.~{Palmeri}, N.~{Rouze}, and K.~{Nightingale},
  ``Imaging transverse isotropic properties of muscle by monitoring acoustic
  radiation force induced shear waves using a 2-d matrix ultrasound array,''
  {\em IEEE Transactions on Medical Imaging}, vol.~32, no.~9, pp.~1671--1684,
  2013.

\bibitem{Schmidt_2016}
J.~L. Schmidt, D.~J. Tweten, A.~N. Benegal, C.~H. Walker, T.~E. Portnoi, R.~J.
  Okamoto, J.~R. Garbow, and P.~V. Bayly, ``Magnetic resonance elastography of
  slow and fast shear waves illuminates differences in shear and tensile moduli
  in anisotropic tissue,'' {\em Journal of biomechanics}, vol.~49, no.~7,
  pp.~1042--1049, 2016.

\bibitem{Chatelin_2016}
S.~Chatelin, I.~Charpentier, N.~Corbin, L.~Meylheuc, and J.~Vappou, ``An
  automatic differentiation-based gradient method for inversion of the shear
  wave equation in magnetic resonance elastography: specific application in
  fibrous soft tissues,'' {\em Physics in Medicine and Biology}, vol.~61,
  no.~13, pp.~5000--5019, 2016.

\bibitem{Singh_2016}
M.~Singh, J.~Li, Z.~Han, C.~Wu, S.~Aglyamov, M.~Twa, and K.~Larin,
  ``Investigating elastic anisotropy of the porcine cornea as a function of
  intraocular pressure with optical coherence elastography,'' {\em Journal of
  Refractive Surgery}, vol.~32, pp.~562--567, 2016.

\bibitem{Singh_2019}
M.~Singh, S.~Wang, C.-H. Liu, J.~Leach, J.~Rippy, I.~V. Larina, J.~F. Martin,
  and K.~V. Larin, {\em Optical coherence elastography reveals the changes in
  cardiac tissue biomechanical properties after myocardial infarction in a
  mouse model}, vol.~10867 of {\em SPIE BiOS}.
\newblock SPIE, 2019.

\bibitem{Parker_2017}
K.~J. Parker, J.~Ormachea, F.~Zvietcovich, and B.~Castaneda, ``Reverberant
  shear wave fields and estimation of tissue properties,'' {\em Physics in
  Medicine and Biology}, vol.~62, no.~3, pp.~1046--1061, 2017.

\bibitem{Ormachea_2018}
J.~Ormachea, B.~Castaneda, and K.~J. Parker, ``Shear wave speed estimation
  using reverberant shear wave fields: Implementation and feasibility
  studies,'' {\em Ultrasound in Medicine and Biology}, vol.~44, no.~5,
  pp.~963--977, 2018.

\bibitem{Ormachea_2019}
J.~Ormachea, K.~J. Parker, and R.~G. Barr, ``An initial study of complete 2{D}
  shear wave dispersion images using a reverberant shear wave field,'' {\em
  Physics in Medicine {\&} Biology}, vol.~64, no.~14, p.~145009, 2019.

\bibitem{Zvietcovich_2019}
F.~Zvietcovich, P.~Pongchalee, P.~Meemon, J.~P. Rolland, and K.~J. Parker,
  ``Reverberant 3d optical coherence elastography maps the elasticity of
  individual corneal layers,'' {\em Nature Communications}, vol.~10, no.~1,
  p.~4895, 2019.

\bibitem{Catheline_2008}
S.~Catheline, N.~Benech, J.~Brum, and C.~Negreira, ``Time reversal of elastic
  waves in soft solids,'' {\em Phys. Rev. Lett.}, vol.~100, p.~064301, Feb
  2008.

\bibitem{Brum_2008}
J.~Brum, S.~Catheline, N.~Benech, and C.~Negreira, ``Shear elasticity
  estimation from surface wave: The time reversal approach,'' {\em The Journal
  of the Acoustical Society of America}, vol.~124, no.~6, pp.~3377--3380, 2008.

\bibitem{Gallot_2011}
T.~{Gallot}, S.~{Catheline}, P.~{Roux}, J.~{Brum}, N.~{Benech}, and
  C.~{Negreira}, ``Passive elastography: shear-wave tomography from
  physiological-noise correlation in soft tissues,'' {\em IEEE Transactions on
  Ultrasonics, Ferroelectrics, and Frequency Control}, vol.~58, no.~6,
  pp.~1122--1126, 2011.

\bibitem{Benech_2013}
N.~Benech, J.~Brum, S.~Catheline, T.~Gallot, and C.~Negreira, ``Near-field
  effects in green's function retrieval from cross-correlation of elastic
  fields: Experimental study with application to elastography,'' {\em The
  Journal of the Acoustical Society of America}, vol.~133, no.~5,
  pp.~2755--2766, 2013.

\bibitem{Jenkins:book}
F.~A. Jenkins and H.~E. White, {\em Fundamental of Optics}.
\newblock McGraw-Hill, $3^{rd}$~ed., 1957.
\newblock {Chapter} 26.

\bibitem{Born:book}
M.~Born and E.~Wolf, {\em Principles in Optics}.
\newblock Pergamon Press, $6^{th}$~ed., 1985.
\newblock {Chapter} 14.

\bibitem{Graff:book}
K.~E. Graff, {\em Wave Motion in Elastic Solids}.
\newblock Dover Publications, 1975.
\newblock Appendix A.

\bibitem{Brum_2014}
J.~Brum, M.~Bernal, J.~L. Gennisson, and M.~Tanter, ``In vivoevaluation of the
  elastic anisotropy of the human achilles tendon using shear wave dispersion
  analysis,'' {\em Physics in Medicine and Biology}, vol.~59, no.~3,
  pp.~505--523, 2014.

\bibitem{Aubry2013}
S.~Aubry, J.~R. Risson, A.~Kastler, B.~Barbier-Brion, G.~Siliman, M.~Runge, and
  B.~Kastler, ``Biomechanical properties of the calcaneal tendon in vivo
  assessed by transient shear wave elastography,'' {\em Skeletal Radiology},
  vol.~42, no.~8, pp.~1143--1150, 2013.

\bibitem{PINSKY_2005}
P.~M. Pinsky, D.~van~der Heide, and D.~Chernyak, ``Computational modeling of
  mechanical anisotropy in the cornea and sclera,'' {\em Journal of Cataract \&
  Refractive Surgery}, vol.~31, no.~1, pp.~136 -- 145, 2005.

\bibitem{Itskov_2002}
M.~Itskov and N.~Aksel, ``Elastic constants and their admissible values for
  incompressible and slightly compressible anisotropic materials,'' {\em Acta
  Mechanica}, vol.~157, pp.~81--96, 2002.

\bibitem{Zvietcovich_2020}
Z.~Fernando, {\em Dynamic optical coherence elastography}.
\newblock Thesis, 2020.

\bibitem{Koo_2013}
T.~K. Koo, J.-Y. Guo, J.~H. Cohen, and K.~J. Parker, ``Relationship between
  shear elastic modulus and passive muscle force: An ex-vivo study,'' {\em
  Journal of Biomechanics}, vol.~46, no.~12, pp.~2053--2059, 2013.

\end{thebibliography}

\end{document}